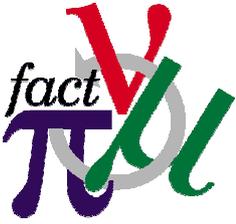
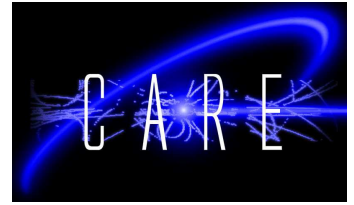

# Future neutrino oscillation facilities: physics priorities and open issues

Alain Blondel

*DPNC Section de Physique 24 quai Ansermet University of Geneva CH 1211 Genève 23 Switzerland*
*based on an Invited plenary presentation at NUFACT05, Frascati, June 2005*

**Abstract:** The recent discovery that neutrinos have masses opens a wide new field of experimentation, for which one has to be ambitious, ingenious… and patient. Accelerator-made neutrinos are essential in this program. Ideas for future facilities include high intensity muon neutrino beams from pion decay ('Superbeam'), electron neutrino beams from nuclei decays ('Beta-beam'), or muon and electron neutrino beams from muon decay ('Neutrino Factory'), each associated with one or several options for detector systems. There are different opinions in the community on how to proceed through this rich choice of questions and possibilities. We now begin a "scoping study" aimed at determining a set of key R&D projects enabling the community to propose an ambitious accelerator neutrino program at the turn of this decade. As an introduction to this study, a set of physics priorities, a summary of the perceived virtues and shortcomings of the various options, and a number of open questions are presented.

## 1   Introduction

### 1.1   Status of the field

The observation of neutrino oscillations has now established beyond doubt that neutrinos have mass and mix. This existence of neutrino masses is in fact the first solid experimental fact requiring physics beyond the Standard Model.

The present status of the field is as follows. Since 1989, we know from LEP [LEPEW] that there are only three families of active light neutrinos coupling to the weak interaction. Since the early 1970's we have hints from solar neutrino experiments that electron neutrinos produced in the sun undergo disappearance on their way to earth, and, from the different disappearance rates measured in different neutrino energy ranges (Chlorine[Homestake], Gallium[GNO],[Sage] Water Cherenkov[Super-Ksolar]) we have indication that matter effects in the sun play an important role [MSW]. This "solar neutrino puzzle" was closed in 2001-2002 with the results from the SNO experiment [SNO01][SNO02], which allowed, by virtue of using heavy water as target, simultaneous measurements of i) the neutral current reaction, providing the total flux of active neutrinos that agreed with solar neutrino flux calculations, and ii) the charged current reaction, providing a measurement of the electron neutrino component representing less than half of the total flux, in agreement with previous observations. The KamLAND [KamLAND02, KamLAND04] experiment provided at the same time a measurement of disappearance of electron anti-neutrinos from nuclear (fission) reactors in Japan, providing, in combination with the solar neutrino results, a precise determination of the relevant neutrino mixing angle of around $30^0$ and of the corresponding mass difference -- that can be expressed as an oscillation quarter-wavelength of L/E ~15000 km/GeV.

Since the late 80's there has been indication from atmospheric neutrino experiments that the muon neutrinos undergo disappearance when going through the earth; this was unambiguously demonstrated by the SuperKamiokaNDE experiment in 1998 [Atmos]. This disappearance takes place at a much shorter quarter-wavelength than for solar neutrinos (L/E~500km/GeV); it is not seen for electron neutrinos, a fact that has been best established by the CHOOZ [CHOOZ] reactor experiment. Recently, it has been confirmed by the K2K experiment in Japan [K2K], the first accelerator neutrino long baseline experiment designed since neutrino masses have been established, and a prototype for future ones.

The above experimental observations are consistently described by three family oscillations, with mass eigenstates {$\nu_1$, $\nu_2$, $\nu_3$} related to the flavour eigenstates {$\nu_e$, $\nu_\mu$, $\nu_\tau$} by a set of Euler angles $\theta_{12}$, $\theta_{13}$, $\theta_{23}$ as depicted in Figure 1. Two independent mass splittings characterize the system, since oscillations only depend on the difference of squared masses. Although no formally agreed definition exists, the usage is that the mass eigenstates are classified by decreasing electron-neutrino content: $|<\nu_e|\nu_1>|^2 > |<\nu_e|\nu_2>|^2 > |<\nu_e|\nu_3>|^2$. With this definition, the mass of $\nu_1$ is not necessarily smaller than that of $\nu_2$. Since neutrino oscillations in vacuum depend on the mass difference as $\sin^2(1.27 \Delta m^2 L/E)$ one cannot determine the sign of $\Delta m^2$ unless the oscillation interferes with another process. In the case of electron neutrinos, this is offered by coherent scattering on electrons in matter, a.k.a. matter effects. The fact that solar neutrinos undergo matter effects in the sun allows us to conclude that know that $\Delta m^2_{12} \equiv \delta m \equiv m^2_2 - m^2_1 > 0$. Since atmospheric neutrino disappearance has only been observed for muon neutrinos, which couple weakly to electron neutrinos at the relevant wavelength, we cannot (yet) tell the sign of the mass difference $\Delta m^2_{13}$, or $\Delta m^2_{23}$, (or $\Delta m^2 \equiv (\Delta m^2_{13} + \Delta m^2_{23})/2$ as suggested by Lisi [Lisi]). The present values of oscillation parameters are summarized in Table 1 and Figure 1.

Neutrino oscillation experiments have demonstrated that neutrinos have masses, but they do not allow a direct measurement of the neutrino masses themselves; only of the differences of squares are accesible. Two complementary methods have been suggested to determine the absolute mass scale. The first is the direct kinematical mass determination in beta decay (usually Tritium, as this isotope has the smallest available Q-value). This has led to limits of about 2 eV/$c^2$ for the mass of the anti-electron-neutrino [Mainz], [Troitsk]. The KATRIN experiment at Karlsruhe is planning to improve the measurements by one order of magnitude [KATRIN] and similar sensitivity could be achieved using Rhenium decay [MARE]. In principle direct measurement of the mass would give three distinct results, $m_{\nu 1}$, $m_{\nu 2}$, $m_{\nu 3}$ with probabilities $|U_{e1}|^2$, $|U_{e2}|^2$, $|U_{e3}|^2$. Measuring the average mass of the neutrino produced in this reaction one would find this:

$$m_{\bar{\nu}_e} = \left( \sum_i |U^2_{ei}| \, m^2_i \right)^{1/2} = \left( \cos^2\theta_{13}(m^2_1 \cos^2\theta_{12} + m^2_2 \sin^2\theta_{12}) + m^2_3 \sin^2\theta_{13} \right)^{1/2}$$

or a similar result depending on the effective sampling of masses of the measurement.

The second set of observables sensitive to neutrino masses comes from astronomy. Within the last ten years, astronomical observations related to the rate of expansion of the early universe and of its large scale structure visible in the power spectrum of the cosmic microwave background have led to a cosmological model, in which the neutrino mass plays an important role; analysis of the existing data leads to limits on the neutrino mass at the level of 0.2-1 eV/$c^2$ (see a more complete discussion in the 2005 BENE report [BENE05] and references therein).

An important question arises when discussing massive neutrinos: is leptonic number conserved? At a very basic level, we are accustomed to fermion number conservation, in the sense that an electron cannot transform itself in a positron, although this would be perfectly allowed kinematically. The same comment is valid for u-type or d-type quarks. Clearly for these charged fermions, charge conservation can be invoked to explain the absence of such transitions. For massless neutrinos of definite helicity (left-handed neutrino and right handed anti-neutrino) we can rely on the conservation of angular momentum to exclude neutrino-anti-neutrino transitions. For massive neutrinos this is no longer the case – the Lorentz-invariant chirality states are no longer identical to the well defined angular momentum states – and neutrino-anti-neutrino transitions at a very small level could occur and violate fermion number, unless nature explicitly forbids it. Thus, unless a new conservation law exists for which no evidence has otherwise been found, massive neutrinos naturally lead to matter-antimatter transitions, a rather significant revolution in the way of thinking of the Standard Model. This, coupled with CP violation is the key to the now popular phenomenon known as Leptogenesis, wherein the matter-antimatter asymmetry of the universe is explained by lepton number violation in the heavy neutrino states ( $\mathbf{N_R}$ and $\mathbf{\overline{N}_L}$ ) and their decays. The generation of baryon asymmetry via leptogenesis from such Majorana Neutrinos [Leptogenesis] would work provided that $m_{\nu i} < 0.1 eV$, for all species of neutrinos.

Two low energy observations would support this attractive construction. The first one would be the observation of neutrinoless double-beta decay ($\beta\beta 0\nu$) in suitable isotopes. In addition to demonstrating unambiguously a process in which lepton number is violated, $\beta\beta 0\nu$ is also sensitive to the neutrino mass scale, although in a different way than the direct kinematical method above.

$$|m_{\beta\beta}| = \left|\sum_i U_{ei}^{*2} m_i\right| = \left|\cos^2\theta_{13}(m_1\cos^2\theta_{12} + m_2 e^{2i\alpha}\sin^2\theta_{12}) + m_3 e^{2i\beta}\sin^2\theta_{13}\right|.$$

The process is extremely rare, ($T_{1/2} \geq 10^{25}$ years or so) and requires exquisitely sensitive experiments. In addition, the interpretation of ββ0ν data depends on calculations of the nuclear matrix elements entering in the process. A claim for a neutrinoless double-β signal has been made by [HM-KK] analyzing the Heidelberg-Moscow data on $^{76}$Ge, $T_{0\nu}$= 1.19 · $10^{25}$ years corresponding to < $m_{\beta\beta}$ >= 0.05 – 0.85 eV (95%CL), the uncertainty coming mainly from the choice of the nuclear matrix element calculation. This result is in contrast with negative results from the same experiment [IGEX1],[IGEX2] and from other experiments [NEMO3] [Cuoricino]. Clearly this is a field in which many experiments are and should be planned (i.e. Cuore, NEMO, EXO, Fiorini, Majorana…) in order to reach the sensitivity to the masses implied by the mass differences observed in neutrino oscillations.

The second would be the observation of leptonic CP violation in neutrino oscillations, and will be extensively discussed in the following.

*Table 1 Neutrino oscillation parameters as of NUFACT05 [Lisi][1]*

| | | |
|---|---|---|
| 'solar parameters' | $\delta m^2 = \pm(7.92 \pm 0.72)\,10^{-5}\,eV^2$ | $\sin^2\theta_{12} = 0.314^{+0.030}_{-0.025}$ |
| 'atmospheric parameters' | $\Delta m^2 = \pm(2.4^{+0.5}_{-0.6})\,10^{-3}\,eV^2$ | $\sin^2\theta_{23} = 0.44^{+0.18}_{-0.10}$ |
| 'solar-atmospheric transition parameters' | | $\sin^2\theta_{13} < 3.2\,10^{-2}$ @ 95% C.L. δ unknown |
| 'absolute mass' | $m_\nu \leq 2.2$ eV (beta-decay)<br>$\Sigma m_\nu \leq O(1\text{ eV})$ (cosmology) | |

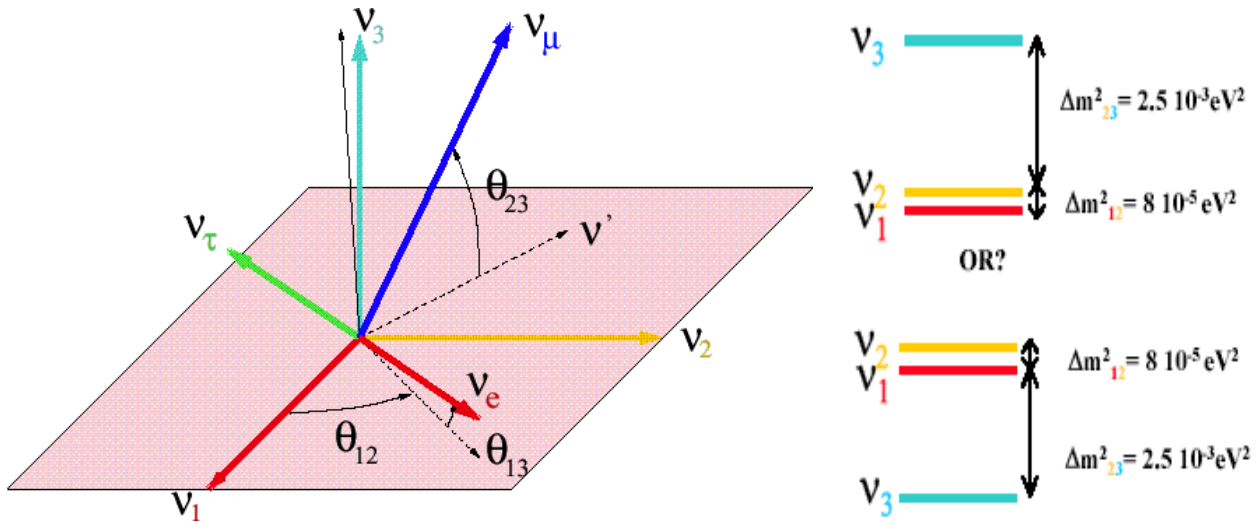

*Figure 1 Present knowledge of the neutrino mixing matrix. The best values are, for the angles, $\theta_{12}$=32⁰, $\theta_{23}$= 45⁰, $\theta_{13}$< 13⁰, and for the masses $\Delta m^2_{12}$ = + 8 $10^{-5}$ $eV^2$, $\Delta m^2_{23}$ = ±2.5 $eV^2$. The unknown phase δ would, if non-vanishing, generate CP and T violation in neutrino oscillations.*

Before leaving this section on the status of the field, it is worth remembering the possible observation of a $\nu_e$ appearance in the proton dump experiment at Los Alamos in the LSND experiment [LSND]. This puzzling observation will be soon verified or falsified by the MiniBoone experiment at Fermilab [MiniBoone]. If it is true this will lead to even more exciting phenomenology!

---

[1] Lisi gives errors as factors and as 95% C.L. ('2 sigma'). I have taken the liberty to express these in a more conventional manner.

## 1.2 Three family oscillations and CP or T violation

It was soon realized that with three families and a favourable set of parameters, it would be possible to observe violation of CP or T symmetries in neutrino oscillations [Ruj99]. This observation reinforced the considerable interest for precision measurements of neutrino oscillation parameters. We know since 2002 and the results from SNO [SNO02] and KamLAND [Kamland02] that the neutrino parameters belong to the so-called LMA solution which suggests that leptonic CP violation should be large enough to be observed in high-energy neutrino oscillation appearance experiments. This has led to extensive studies, such as those published recently in the CERN yellow report [ECFAreport], or in a recent BENE [BENE] workshop on physics at a high intensity proton driver [MMW].

The phenomenon of CP (or T) violation in neutrino oscillations manifests itself by a difference in the oscillation probabilities of say, $P(\nu_\mu \to \nu_e)$ vs $P(\bar{\nu}_\mu \to \bar{\nu}_e)$ (CP violation), or $P(\nu_\mu \to \nu_e)$ vs $P(\nu_e \to \nu_\mu)$ (time reversal violation). It can be noticed right away that observation of this important phenomenon requires appearance experiments; indeed a reactor or solar neutrino experiment, sensitive to the disappearance $P(\nu_e \to \nu_e)$ which is clearly time-reversal invariant, would be completely insensitive to it. This can be seen as an advantage in view of a precise and unambiguous measurement of the mixing angles; for the long term goal of observing and studying CP violation, we are confined to appearance experiments. The $\nu_\mu \to \nu_e$ transition can be expressed as

$$
\begin{aligned}
p(\nu_\mu \to \nu_e) = \\
& 4 c_{13}^2 s_{13}^2 s_{23}^2 \sin^2 \frac{\Delta m_{13}^2 L}{4E} \quad &\theta_{13} \text{ driven} \\
& + 8 c_{13}^2 s_{12} s_{13} s_{23} (c_{12} c_{23} \cos\delta - s_{12} s_{13} s_{23}) \cos \frac{\Delta m_{23}^2 L}{4E} \sin \frac{\Delta m_{13}^2 L}{4E} \sin \frac{\Delta m_{12}^2 L}{4E} \quad &CP-even \\
& - 8 c_{13}^2 c_{12} c_{23} s_{12} s_{13} s_{23} \sin\delta \sin \frac{\Delta m_{23}^2 L}{4E} \sin \frac{\Delta m_{13}^2 L}{4E} \sin \frac{\Delta m_{12}^2 L}{4E} \quad &CP-odd \\
& + 4 s_{12}^2 c_{13}^2 \{c_{12}^2 c_{23}^2 + s_{12}^2 s_{23}^2 s_{13}^2 - 2 c_{12} c_{23} s_{12} s_{23} s_{13} \cos\delta\} \sin \frac{\Delta m_{12}^2 L}{4E} \quad &\text{solar driven} \\
& - 8 c_{13}^2 s_{13}^2 s_{23}^2 \cos \frac{\Delta m_{23}^2 L}{4E} \sin \frac{\Delta m_{13}^2 L}{4E} \frac{aL}{4E}(1-2s_{13}^2) \quad &\text{matter effect (CP odd)}
\end{aligned}
$$
(1)

*Figure 2: complete formula for the $\nu_\mu \to \nu_e$ transition; $c_{12}$ stands for $\cos\theta_{12}$, etc... and $\hbar c=1$.*

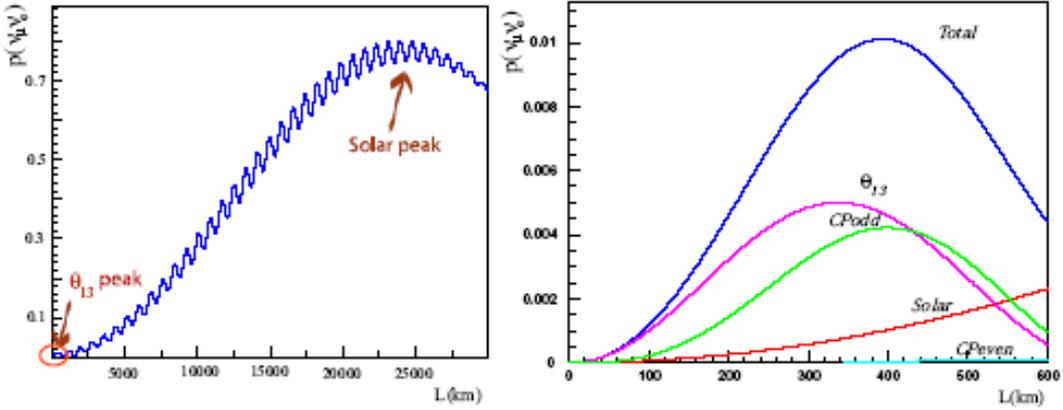

*Figure 3 Description of neutrino oscillations for 1 GeV neutrinos as a function of distance to the source. The oscillation parameters are as in Table 1, with : $\sin^2 2\theta_{13} = 0.01$, $\delta = 0$ for the left plot, $\delta = -\pi/2$ for the one on the right.*

One of the interesting aspects of this formula is the occurrence of matter effects which, unlike the straightforward $\theta_{13}$ term, depend on the sign of the mass difference $\Delta m^2_{13}$. These terms should allow extraction of the mass hierarchy, but could also be seen as a background to the CP violating effect, from which they can be distinguished by the very different neutrino energy dependence, matter effects being larger for higher energies, with a 'matter resonance' at about 12 GeV.

The CP violation can be seen as interference between the solar and atmospheric oscillation for the same transition, as emphasized in Figure 3. Of experimental interest is the CP-violating asymmetry $A_{CP}$:

$$A_{CP} = \frac{P(\nu_e \to \nu_\mu) - P(\overline{\nu}_\mu \to \overline{\nu}_e)}{P(\nu_e \to \nu_\mu) + P(\overline{\nu}_\mu \to \overline{\nu}_e)} \approx \frac{\sin\delta \; \sin\vartheta_{13} \; \sin(\Delta m_{12}^2 \, L/4E) \; \sin\vartheta_{12}}{\sin^2\vartheta_{13} + \textbf{solar term}}$$

or the equivalent time reversal asymmetry $A_T$ which is displayed on Figure 4.

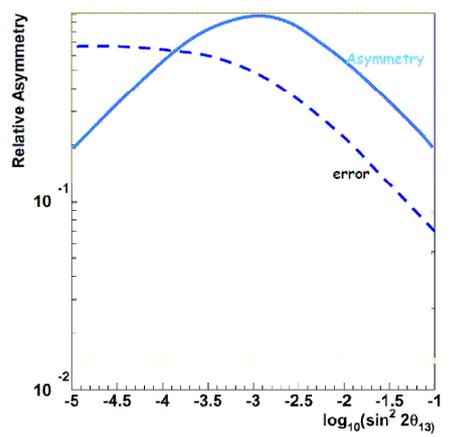

*Figure 4 Magnitude of the time reversal asymmetry at the first oscillation maximum, for $\delta=1$ as a function of the mixing angle $\sin^2 2\theta_{13}$. The curve marked 'error' indicates the $\theta_{13}$ dependence of the statistical error on such a measurement for a given flux and detector mass.*

The asymmetry can be large and its value increases for smaller values of $\theta_{13}$ up to the value when the two oscillations (solar and atmospheric) are of the same magnitude. The following remarks can be made:

1. The ratio of the asymmetry to the statistical error is fairly independent on $\theta_{13}$ for large values of this parameter, which explains the relative flatness of the sensitivity curves.

2. This asymmetry is valid for the first maximum. At the second oscillation maximum the curve is shifted to higher values of $\theta_{13}$ so that it could be then an interesting possibility for measuring the CP asymmetry, although the reduction in flux is considerable.

3. The asymmetry has opposite sign for $\nu_e \to \nu_\mu$ and $\nu_e \to \nu_\tau$, and changes sign when going from one oscillation maximum to the next.

4. The asymmetry is small for large values of $\theta_{13}$, placing a challenging emphasis on systematics in the cross-section measurements.

This last point deserves a dedicated discussion. As will be seen later, facilities proposed for large values of $\theta_{13}$ are superbeams and beta-beams of relatively low energy, typically aiming at a detector made of water or scintillator. If the initial beam consists purely of either electron neutrinos or antineutrinos, as it is the case for the beta-beam, of (almost) purely of muon neutrinos and anti-neutrinos, as is the case for a superbeam, it will be difficult in practice to measure precisely the oscillation probability, $P(\nu_\mu \to \nu_e)$ or $P(\nu_e \to \nu_\mu)$, by lack of a precise knowledge of the cross-section for the final state neutrino which is absent in the initial beam and cannot be measured in the near detector. Although some of the related errors can perhaps be reduced in the comparison between neutrinos and antineutrinos, this difficulty could strongly advocate for a facility which combines both types of neutrinos, i.e. the beta-beam+superbeam combination, or the neutrino factory.

## 1.3  Strategy and considered facilities

In order to design a facility it is important to delineate the main physics objective that will drive the choice of parameters, while keeping in mind other important physics outcomes and interesting by-products that will constitute interesting selling points of the facility. Of course such a hierarchy of physics relevance is a matter of choice and is somewhat subjective. A common view needs to be accepted by the community. The following is my own prejudice.
1. Main objective: Observe and study CP and T violation, determine mass hierarchy. This can be done provided neutrino oscillation probabilities are measured with great precision, in an appearance channel involving electrons, and over a broad range of energies to decipher the matter effect from the CP violation.

2. Important objectives: unambiguous precision measurements of mixing angles and mass differences.

3. by-products: precision short baseline neutrino physics, unitarity tests, nuclear physics, muon collider preparation, muon EDM.

4. Other physics capabilities: nucleon decay, observation of cosmic events (supernovae, cosmic ray bursts, etc..), other particle physics (muon lepton flavour violating decays)

Can we make one facility that will do all of this? Or do we prefer an approach where these pieces will be produced one at a time by individual dedicated experiments?

Let me take here a purely European point of view, and quote the conclusions of the SPSC workshop in Villars, *"Future neutrino facilities offer great promise for fundamental discoveries (such as leptonic CP violation) in neutrino physics and a post LHC funding window may exist for a facility to be sited at CERN"*. An ambitious neutrino programme is thus a distinct possibility, but it must be well prepared to have a good proposal in time for the big decision period around 2010, when, LHC results being available, the future of particle physics will be decided to a large extent.

What can we expect the situation to be in 2010 as far as the neutrino oscillation business is concerned? In Figure 5, the evolution of the world combined sensitivity to $\theta_{13}$ is shown as a function of the year. It is clear that by that time our knowledge will have improved by almost one order of magnitude. It is clear that by the end of the current decade (end 2010), we will know whether $\sin^2 2\theta_{13}$ is larger or smaller than 1%, and that we should be able by then to propose a facility accordingly.

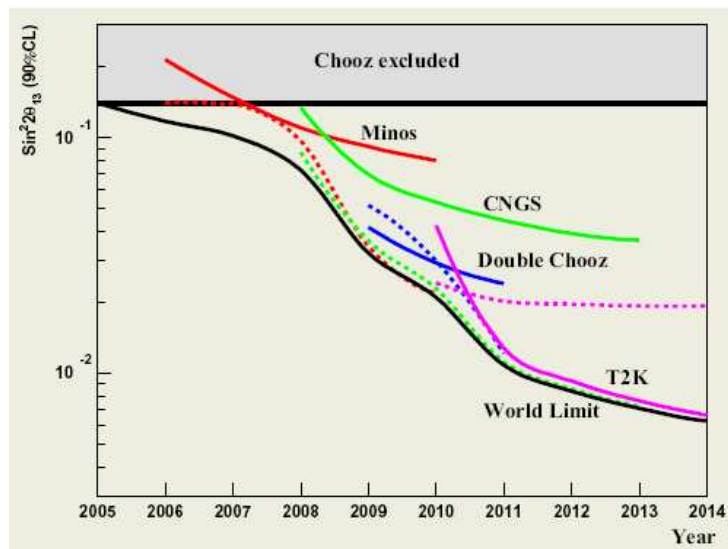

*Figure 5 Sensitivity to $\sin^2 2\theta_{13}$ as function of year taking into account presently approved projects.*

The facilities that have been considered promising for observation of CP violation are as follows

1. Superbeam alone + large detector(s) (e.g. T2HK, NOvA)

2. Superbeam + Beta-Beam + Megaton detector (SB+BB+MD) (e.g. Fréjus)

3. Neutrino Factory (NuFact) + magnetic detector (40kton)

The physics abilities of the neutrino factory have been advocated to be superior, but the question in everyone's mind is « what is the realistic time scale? ». To a large extent the question of time scale should be decided by the cost of the considered facility. The (Hardware) cost estimate for a neutrino factory was estimated in [APS04] (Not found in references) to be ~1B€ + detectors, with rather large uncertainties since the original cost estimate was based on a somewhat different design [studyII]. This estimate needs to be verified and ascertained on a localized scenario (CERN, RAL…) and accounting.

The cost of a (BB+SB+MD) is not very different A large cost driver here (or in the T2HK option) is the very large detector, the cost of which is at the moment quite difficult to estimate, since there will be a hard limit on the size of the largest underground cavern that can be excavated. The issues related to beta-beam are subject to a design study under Eurisol at the moment, and those related to the high power superbeam (4MW on target) are similar to those of a neutrino factory.

From this brief discussion it is very clear that a cost/physics performance/feasibility comparison is needed; this will be the object of the upcoming 'scoping study'.

## 2   Description of the facilities

### 2.1   Off axis superbeams: T2K, T2HK and NovA

The facilities that are envisaged immediately after the CNGS and MINOS-on axis have been designed so that the neutrino beam energy matches the distance to the detector. Since a Wide-band beam on axis has an energy which is approximately 5-10% of the incident proton energy, with a rapidly decreasing event rate as the horn optics are detuned to produce a low energy spectrum, it is more advantageous to tune the beam energy by using the off-axis trick. This is what is being built for the T2K experiment, and envisaged for the off axis NUMI experiment, NOvA.

The T2K (Tokai to Kamioka) experiment will aim neutrinos from the Tokai site to the SuperKamiokaNDE detector 295 km away (Figure 6). The main design features of the T2K experiment lie in its beam line:

The neutrino beam is produced by pion decay from a horn focused beam, with a system of three horns and reflectors. The decay tunnel length (130 m long) is optimised for the decay of 2-8 GeV pions and short enough to minimize the occurrence of muon decays.

 The neutrino beam is situated at an angle of 2-3 degrees from the direction of the super-Kamiokande detector, assuring a pion decay peak energy of 0.6 GeV

 The beam line is equipped with a set of dedicated on-axis and off-axis detectors at two different distances, 280 meters, and possibly, at a later stage, 2 km as shown in Figure 7.

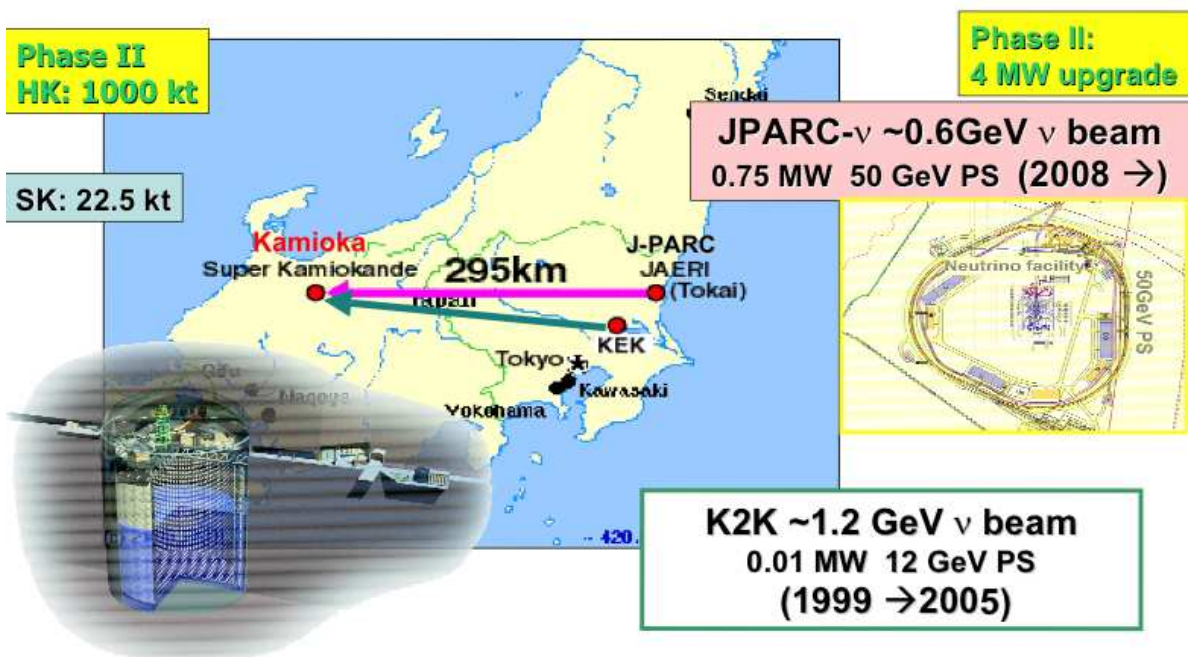

**Figure 6:  Overview of the J-PARC to SuperKamiokande long baseline experiment foreseen from 2009. Also shown is the K2K beam from KEK to the SuperKamiokande detector.**

The main goals of the experiment are as follows:

1. The highest priority goal of the experiment, is the search for the yet unobserved neutrino oscillation $\nu_\mu \leftrightarrow \nu_e$ with the same frequency as the atmospheric oscillation (which is known to be mostly due to $\nu_\mu \leftrightarrow \nu_\tau$). This will manifest itself by appearance of events with an electron in the final state. This reaction is driven by the yet unknown mixing angle $\theta_{13}$, which is also driving the CP violation asymmetry which could be observed in the same channel by comparing neutrino oscillations to antineutrino oscillations. The main challenge here is the understanding of all the background channels that produce or mimic an electromagnetic shower: beam $\nu_e$ from K and muon decay; $\pi^0$ production by neutral current events. For that purpose, a dedicated fine grain detector is under construction for the 280m detector station, where the rate is higher. It will measure precisely the beam composition and the rate of backgrounds, so as to be able to perform a good simulation of the far detector. Since nuclear effects are rather important (pion absorption in particular), the detector material should be as similar as possible to the water of Super-Kamiokande. It is expected that the sensitivity of the experiment will be of the order of $\sin^2 2\theta_{13} \leq 0.006$.

2. Disappearance measurements, where the number and rate of $\nu_\mu \rightarrow \nu_\mu$ events is studied. This will improve measurement of $\Delta m^2_{13}$ down to a precision of a 0.0001 or so. The exact measurement of the maximum disappearance is a precise measurement of $\sin^2 2\theta_{23}$. These precision measurements of already known quantities require good knowledge of flux shape, absolute energy scale, experimental energy resolution and of the cross-section as a function of energy. In this case, it is not only absolutely necessary to have a near detector station made with the same material but also it should have the same acceptance as the far detector. The flux at the 2 km station is much more similar to the SK flux than at 280 m (Figure 7). This constitutes an argument in favour of a 2 km near-detector, for this particular physics goal.

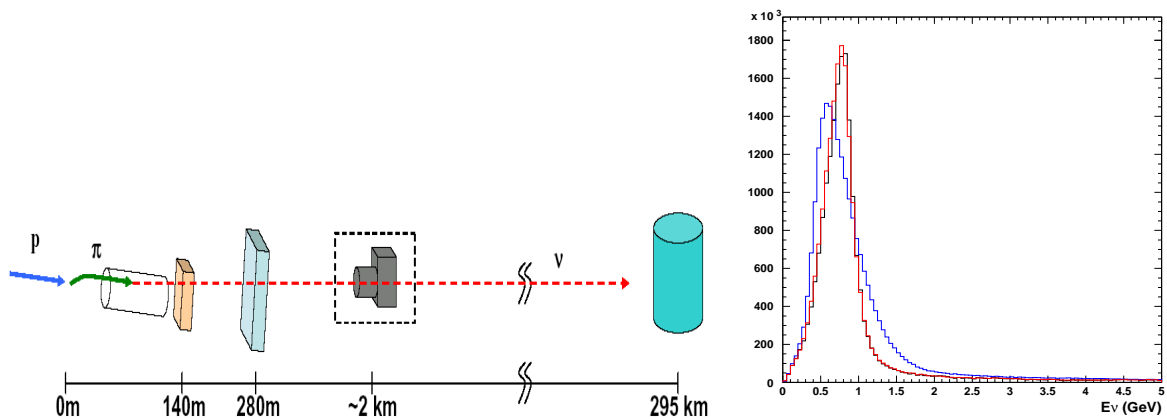

**Figure 7: Schematic description of the detectors along the T2K beam line. The 280m detector station is part of the approved project. The 2km station is part of a possible upgrade. Right: flux shape from the off-axis neutrino beam at 280 m (blue) 2km (red) of Super-Kamiokande (black).**

The T2K experiment is scheduled to start in 2009 with a beam intensity reaching 1 MW beam power on target after a couple years. It has an upgrade path which involves: a 2 km near detector station featuring a water Cherenkov detector, a muon monitor and a fine grain detector (possibly liquid argon); an increase of beam power up to the maximum feasible with the accelerator and target (4 MW?); and a very large water Cherenkov (HyperKamiokande) with a rich physics programme in proton decay, atmospheric and supernova neutrinos and, perhaps, leptonic CP violation, that could be built in about 15-20 years from now. The difficulty in the experiment is that it runs at relatively low neutrino energies, so that the $\nu_e$ appearance signal is situated on top of an intrinsic background from the beam and from $\pi^0$ decays.

The Fermilab NUMI beam is exploiting at present about half a MW of beam power and an off axis detector location has been identified. The experiment would be made deliberately complementary to T2K by using higher beam energy (1.5 GeV). At these energies the channels with one or few pions in the final state are open and a Water Cherenkov would suffer too much background. A fully active liquid scintillator detector is being studied, NovA, with capability to separate electrons from charged and neutral pions. The higher energy w.r.t the T2K programme allows to some extent and for favourable values of the parameters, a sensitivity to the matter effect. More information on the performance and outlook for upgrades can be found in the proposal [NovA] and presentations at NUFACT05 [Plunkett].

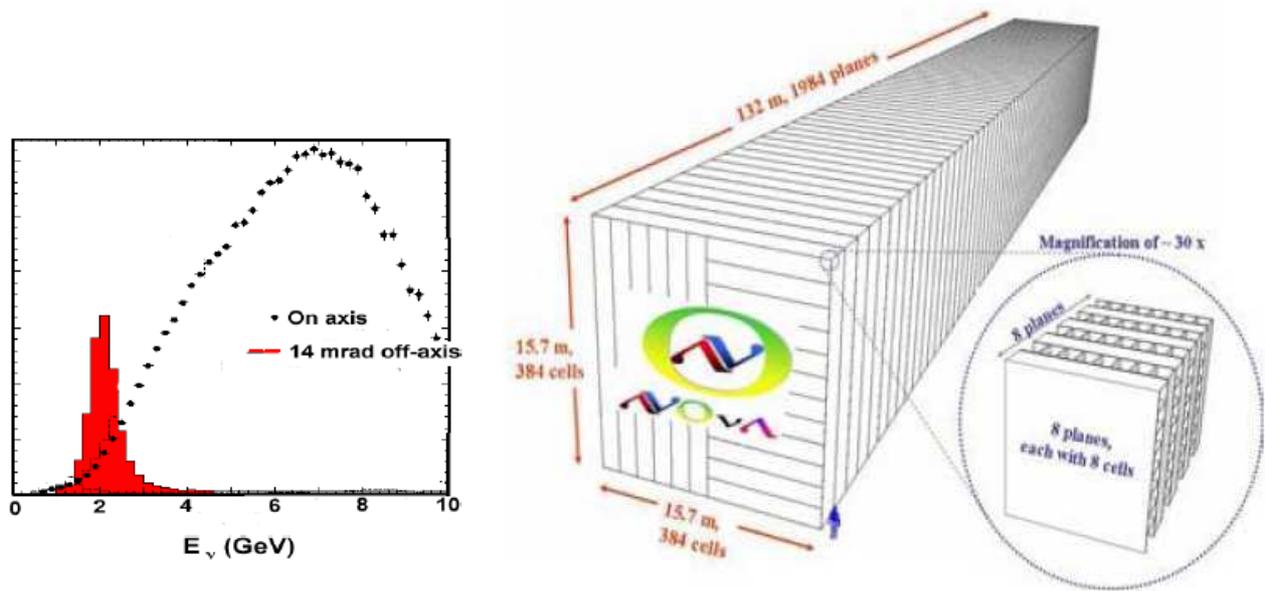

*Figure 8 Left: the 14 mrad off-axis NUMI beam; right: the considered 20 kton fully active scintillator NovA detector.*

## 2.2 The beta-beam + Superbeam facility

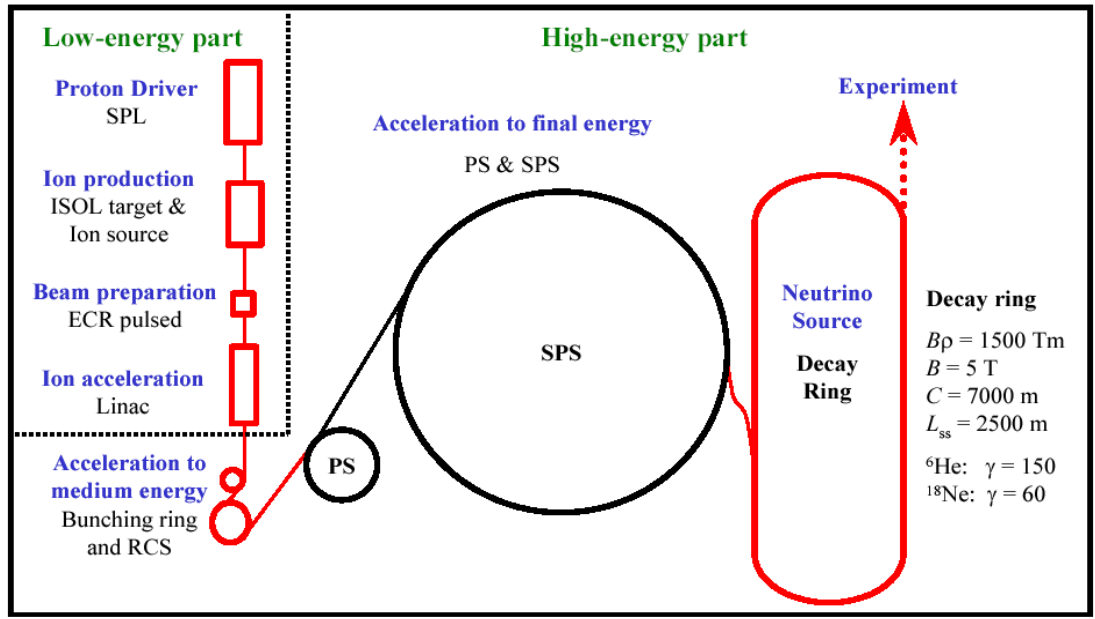

*Figure 9 Beta-beam base line design, partially using existing CERN accelerator infrastructure (parts in black).*

The beta-beam concept [Zucchelli] is based on the acceleration, storage and beta-decay of suitable ions. The preferred ions are

$^6\text{He}^{++} \rightarrow {}^6\text{Li}^{+++} e^- \bar{\nu}_e$

$^{18}\text{Ne} \rightarrow {}^{18}\text{F } e^+ \nu_e$

or $^{150}\text{Dy} + e^- \rightarrow {}^{150}\text{Tb } \nu_e$

The first one is normal beta decay and produces a pure wide-band flux of electron anti-neutrinos. The second is the beta-plus decay and produces a pure electron neutrino beam. The third one, electron capture on heavier nuclei, is a relatively newer idea by Bernabeu [Bernabeu04, Burget05] which would allow producing a pure, monochromatic, electron neutrino beam.

The great interest of the beta-beam lies in its purity, and its relative practicality [beta-beam]: as long as one contents oneself with existing proton machines, the additional required infrastructure is limited to a (quite challenging) high intensity ion source, and a storage ring. The main drawback is that this leads to relatively low energy neutrinos $E_\nu = 2 \gamma E_0$ where $E_0 \sim 3$ MeV is the energy of the neutrino in the decay at rest and $\gamma$ is the Lorentz boost of the accelerated ion. At the CERN SPS, one can accelerate protons to 450 GeV, thus $^6$He to 150 GeV/u or $\gamma < 150$. This limits the neutrino energy to about 600 MeV, while already requiring construction of a storage ring with a rigidity equivalent to that of the SPS. Similarly to the superbeam case, the detector of choice for a low energy beta-beam is a large water Cherenkov. For higher energies the technology would change similarly to a fine grained detector, using scintillator or liquid argon. The higher cross-section and natural focusing at high energy compensates the more difficult realization of massive segmented detectors.

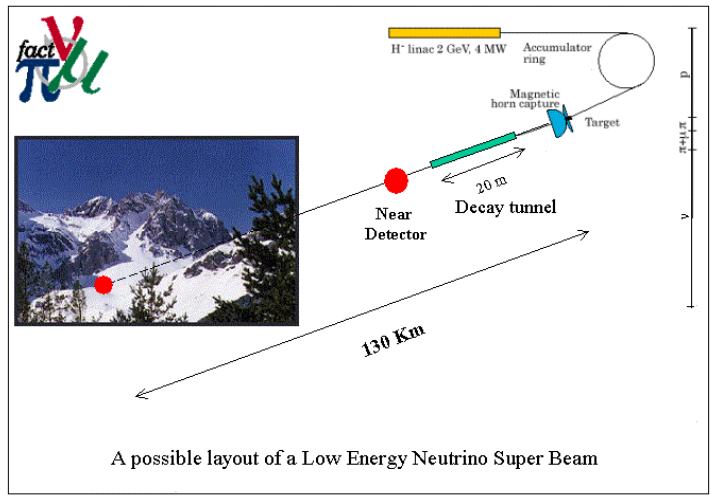

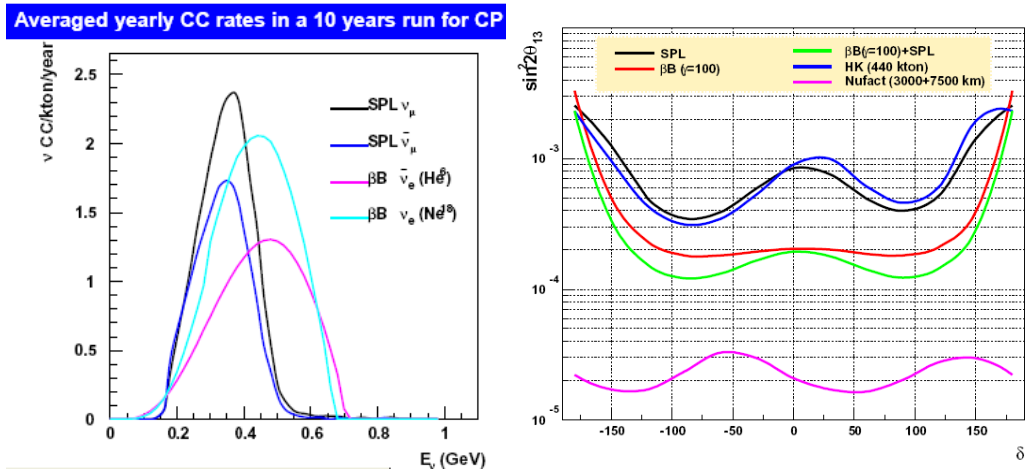

*Figure 10 The beta-beam + superbeam + megaton facility. Top: the schematic layout; bottom left: the (non-oscillated) event numbers for a run of 2 years of neutrinos and 8 years of anti-neutrinos; right the sensitivity to $\sin^2 2\theta_{13}$ as a function of the phase $\delta$ for this set-up in comparison with T2HK and the neutrino factory.*

The high intensity flux seems reasonably easy to obtain for anti-neutrinos with the $^6$He, but $^{18}$Ne appears to be more difficult, perhaps smaller by one order of magnitude. The production of Dy seems even more limited; the application may be a wonderful way to measure cross-sections and nuclear effects directly with a monochromatic beam in the near detector.

The superbeam would be a standard horn-focused neutrino beam from pion decay, produced from low energy protons, with the advantage that the limited kaon production leads to a small and controllable component of electron neutrinos in the beam, from muon decays. This can actually be varied and monitored by varying the length of the decay tunnel.

There exists a "baseline scenario" at CERN for a superbeam + beta beam facility pointing at a megaton water Cherenkov in the Fréjus laboratory, with a baseline of 130 km. As described by J.E. Campagne [Campagne05], the superbeam can be improved by increasing the proton beam energy to 3.5 GeV over the original SPL at 2.2 GeV (Figure 10 )

It has been pointed out that beta-beam at higher energies would be more powerful, especially since the number of events per ton scales as $\gamma^3$, as shown in Figure 11. It has been argued that a beta-beam could be run at the Tevatron with a higher energy, $\gamma \sim 350$ for the Helium beam, and 580 for the Neon. Clearly the cost of such a facility increases rapidly with energy, since a storage ring of equivalent rigidity would have to be constructed. This should be taken into account when imagining high energy beta beams. In addition the background issues do increase when energy increases, since pion production (the main background to the muon signal) increases rapidly with neutrino energy.

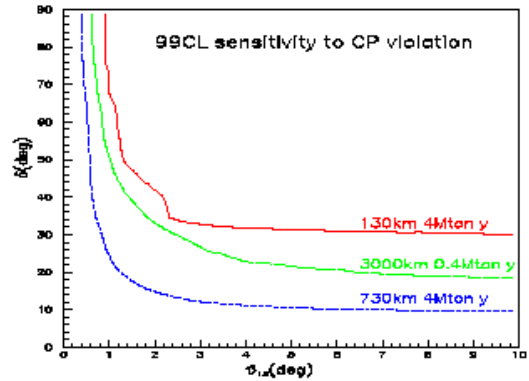

*Figure 11 Comparative performance of the baseline beta-beam scenario (red) with a higher energy one ($\gamma= 350$ for $^6Ne$, as could be accelerated in a 1TeV proton accelerator) blue line. The same far detector is being assumed (1 Mton water Cherenkov).*

## 2.3 Detectors

The detector for the low energy single flavour beams must be designed for the largest possible mass, while providing good separation between electrons, muons and pions. The water Cherenkov is an obvious choice, with the caveat already mentioned that it works best in the energy region below the effective pion production threshold, where $\pi^0$ s mimic electrons, and $\pi^{+-}$ mimic muons. (600MeV to 1 GeV). In this energy range, the water Cherenkov is unchallenged (Figure 12). At higher energies, more information is needed and liquid scintillator (NovA) and liquid argon detectors have been proposed. A total mass of the order of 30 kton seems feasible for scintillator (Figure 8), and ideas for up to 100 kton are being considered for the liquid argon detectors (Figure 13).

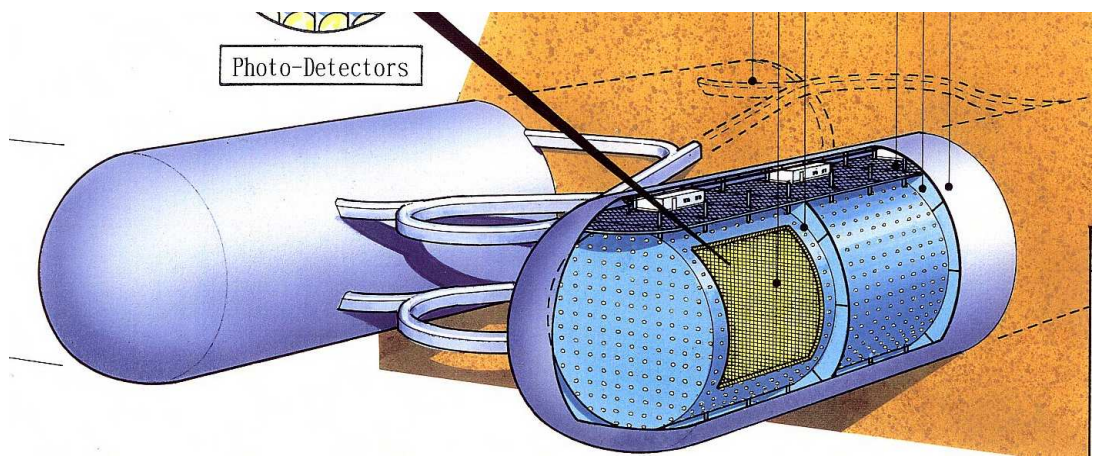

*Figure 12 Sketch of a Mton water Cherenkov: here the Hyper-KamiokaNDE detector. The nearly "cubic" design that could be achieved for the Super-KamiokaNDE detector is no longer feasible due to limitations in underground excavations. One of the dimensions needs to be expanded (here the length).*

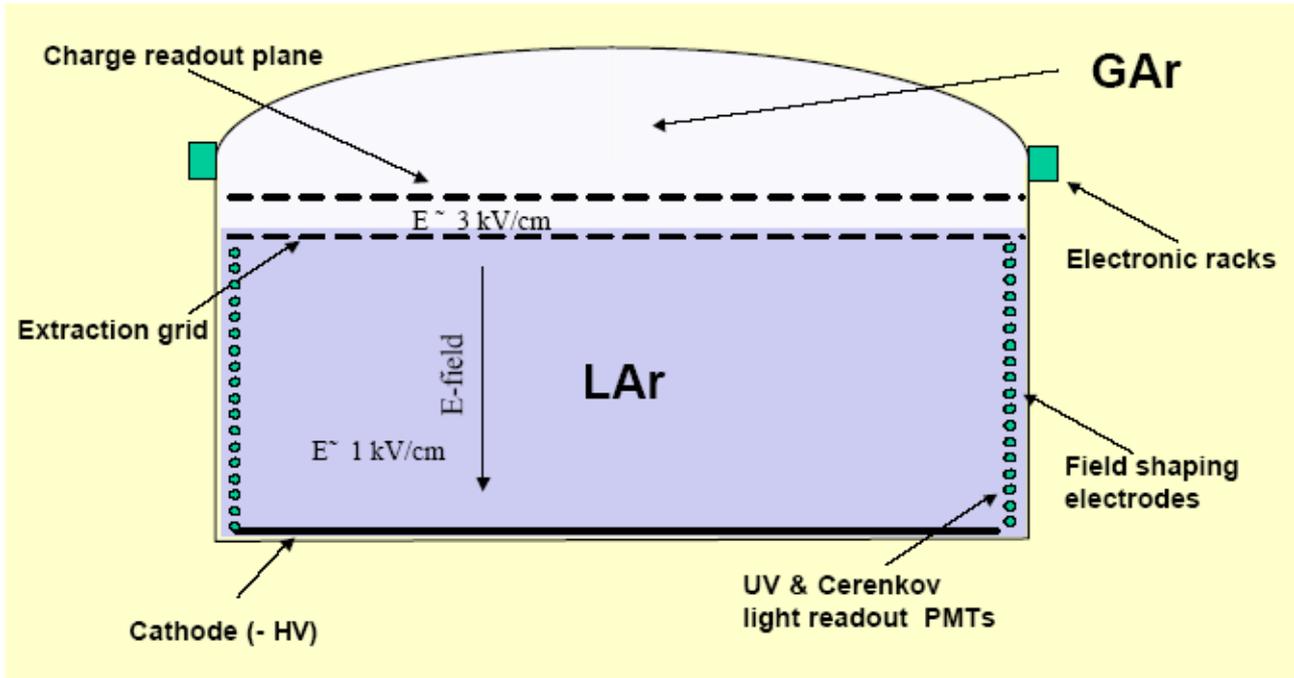

Figure 13: Schematic layout of the inner detector of a future Large Liquid Argon detector [Rubbia]

## 2.4 The neutrino factory

In a Neutrino Factory [nufact] muons are accelerated from an intense source to energies of several GeV, and injected in a storage ring with long straight sections. The muon decays:

$$\mu^+ \to e^+ \nu_e \bar{\nu}_\mu \quad \text{and} \quad \mu^- \to e^- \nu_\mu \bar{\nu}_e$$

provide a very well known flux with energies up to the muon energy itself. The overall layout is shown in Figure 14.

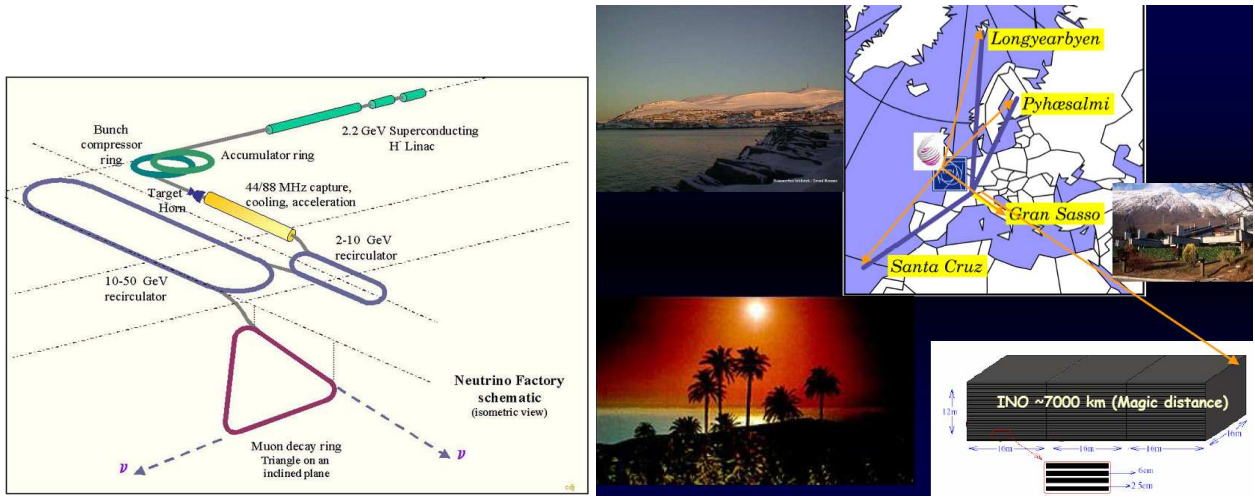

Figure 14 Left: Schematic layout of a Neutrino Factory; right: possible long baseline scenarios for a European based facility.

Neutrino Factory designs have been proposed in Europe [Aut99], [Gru02], the US [MuColl] [StudyI][StudyII], and Japan [Japnufact]. Of the three designs, the one in the US is the most developed, and we will use it as a example in general with a few exceptions. The conclusion of these studies is that, provided sufficient resources, an accelerator complex capable of producing and storing more than $10^{21}$ muons per year can be built. The Neutrino Factory consists of the following subsystems:

**Proton Driver.** It provides 1-4 MW of protons on a pion production target. For the Neutrino Factory application the energy of the beam is not critical, since it has been shown that the production of pions is roughly proportional to beam power. The time structure of the proton beam has to be matched with the time spread induced by pion decay (1-2 ns); for a linac driver such as the SPL, this requires an additional accumulator and compressor ring.

**Target, Capture and Decay.** A high-power target sits within a 20T superconducting solenoid, which captures the pions. The high magnetic field smoothly decreases to 1.75T downstream of the target, matching into a long solenoid decay channel. A design with horn collection has been proposed at CERN for the Neutrino Factory, with the benefit that it can be also used for a superbeam design. The advantage of the horn that it sign-selects the pions and muons is compensated by the fact that in a Neutrino Factory design one could accelerate both signs of muons, thus doubling the available flux.

**Bunching and Phase Rotation.** The muons from the decaying pions are bunched using a system of RF cavities with frequencies that vary along the channel. A second series of RF cavities with higher gradients is used to rotate the beam in longitudinal phase-space, reducing the energy spread of the muons.

**Cooling.** A solenoid focusing channel with high-gradient 201 MHz RF cavities and either liquid-hydrogen or LiH absorbers is used to reduce the transverse phase-space occupied by the beam. The muons lose, by ionisation, both longitudinal- and transverse-momentum as they pass through the absorbers. The longitudinal momentum is restored by re-acceleration in the RF cavities.

**Acceleration.** The central momentum of the muons exiting the cooling channel is 220 MeV/c. A superconducting linac with solenoid focusing is used to raise the energy to 1.5 GeV. Thereafter, a Recirculating Linear Accelerator raises the energy to 5 GeV, and a pair of Fixed-Field Alternating Gradient rings accelerates the beam to at least 20 GeV.

**Storage Ring.** A compact racetrack geometry ring is used, in which 35% of the muons decay in the neutrino beam-forming straight section. If both signs are accelerated, one can inject in two superimposed rings or in two parallel straight sections.

This scheme produces over $6 \times 10^{20}$ useful muon decays per operational year and per straight section in a triangular geometry. The European Neutrino Factory design is similar to the US design, but differs in the technologies chosen to implement the subsystems. The Japanese design is very different, and uses very large acceptance accelerators rather than cooling.

An important Neutrino Factory R&D effort is ongoing in Europe, Japan, and the U.S. since a few years. Significant progress has been made towards optimising the design, developing and testing the required components, and reducing the cost. To illustrate this progress, the cost estimate for a recent update of the US design [APS04] is compared in *Table 2* with the corresponding cost for the previous "Study II" US design [Study II]. It should be noted that the Study II design cost was based on a significant amount of engineering input to ensure design feasibility and establish a good cost basis.

Neutrino Factory R&D has reached a critical stage in which support is required for two key international experiments (MICE [MICE] and Targetry [target-exp]) and a third-generation international design study. If this support is forthcoming, a Neutrino Factory could be added to the Neutrino Physics roadmap in less than a decade.

Table 2 Comparison of unloaded Neutrino Factory costs estimates in M$ for the US Study II design and improvement estimated for the latest updated US design. Costs are shown including A: the whole complex; B no Proton Driver; C no proton driver and no Target station in the estimates. Table from Ref. [APS04].

| Costs in M$ | A | B | C |
|---|---|---|---|
| Old estimate from Study II | 1832 | 1641 | 1538 |
| Multiplicative factor for new estimate | 0.67 | 0.63 | 0.60 |

## 2.5 Oscillations physics at a neutrino factory

Considering a Neutrino Factory with simultaneous beams of positive and negative muons, the following 12 oscillation processes can be studied.

| $\mu^+ \to e^+ \nu_e \bar{\nu}_\mu$ | $\mu^- \to e^- \nu_\mu \bar{\nu}_e$ | |
|---|---|---|
| $\bar{\nu}_\mu \to \bar{\nu}_\mu$ | $\nu_\mu \to \nu_\mu$ | disappearance |
| $\bar{\nu}_\mu \to \bar{\nu}_e$ | $\nu_\mu \to \nu_e$ | appearance ("platinum" channel?) |
| $\bar{\nu}_\mu \to \bar{\nu}_\tau$ | $\nu_\mu \to \nu_\tau$ | appearance (atmospheric oscillation) |
| $\nu_e \to \nu_e$ | $\bar{\nu}_e \to \bar{\nu}_e$ | disappearance |
| $\nu_e \to \nu_\mu$ | $\bar{\nu}_e \to \bar{\nu}_\mu$ | appearance: "golden" channel |
| $\nu_e \to \nu_\tau$ | $\bar{\nu}_e \to \bar{\nu}_\tau$ | appearance: "silver" channel |

An important feature of the Neutrino Factory is the possibility of having opposite muon charges circulating in the ring, therefore allowing also the study of the charged-conjugated processes of those above. Of course the neutrinos coming from decays of muons of different charge must no be confused with each other, this can be done by timing provided the storage ring is adequately designed.

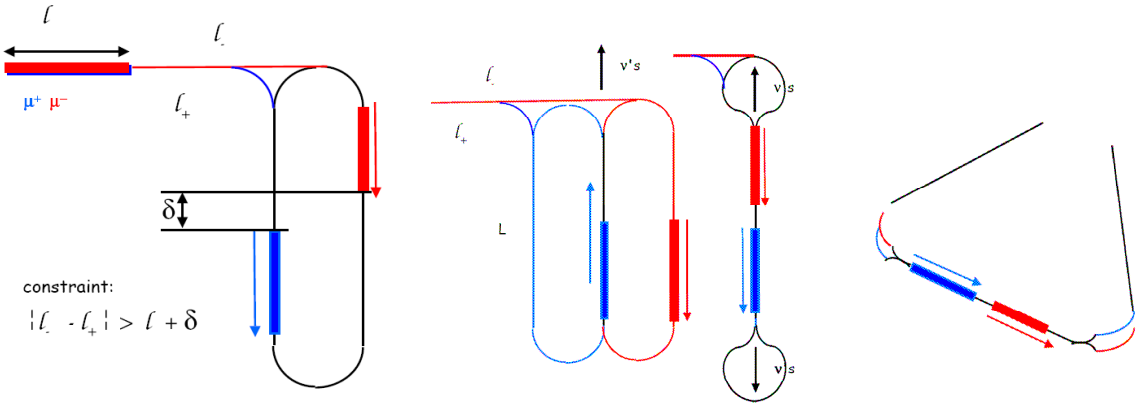

*Figure 15 Neutrino factory with both signs accelerated and stored. Left: in the race-track geometry, the two long straight sections must point toward the same far detector, but two near detector stations are required. It may be more practical to produce both decays in the same straight section, onto the same near detector, as in a double ring or in a dog-bone geometry (middle). For the triangle, one must foresee two superposed rings or special magnets for the bends. [Blondel04]*

It remains to separate the two neutrino flavours that are always simultaneously present. Identification of the flavour of the lepton produced in charged-current interactions is not sufficient and one needs to measure its charge. For muons in the final state (coming from $\nu_\mu$ interactions of from decays of $\tau \to \mu\nu\nu$ ), this can be done readily using a magnetic detector of design similar to that of the CDHS or MINOS experiments, for which by that time one can safely assume that it could be built with a mass of the order of 100 ktons [Nelson]. Many studies have been performed under this hypothesis, where the main discovery channel is the 'wrong sign muon' also called "golden" channel [Cervera00].

Compared with conventional neutrino beams, Neutrino Factories yield higher signal rates with lower background fractions and lower systematic uncertainties. These characteristics enable Neutrino Factory experiments to be sensitive to values of $\theta_{13}$ that are beyond the reach of any other proposed experiment. Several studies (see e.g. [Huber03]) have shown that a non-zero value of $\sin 2\theta_{13}$ could be measured for values as small as $O(10^{-4})$. In addition, both the neutrino mass hierarchy and CP violation in the lepton sector could be measured over this entire range. Even if $\theta_{13} = 0$ the probability for $\nu_e \to \nu_\mu$ oscillations in a long-baseline experiment is finite, and a Neutrino Factory would still make the first observation of it in an appearance experiment, and put a sufficiently stringent limit on the magnitude of $\theta_{13}$ to suggest perhaps the presence of a new conservation law. For the measurement of the quantities $\theta_{13}$ and $\delta$, it has been shown that the golden observables are the oscillation probabilities $\nu_e \to \nu_\mu$ and $\bar{\nu}_e \to \bar{\nu}_\mu$ at baselines, L, and energies, E, in the atmospheric range $E/L \approx \Delta m^2_{23}$. The $\text{sign}(\Delta m^2_{23})$ can be determined with these same transitions given the high energy and long baselines available (Figure 16).

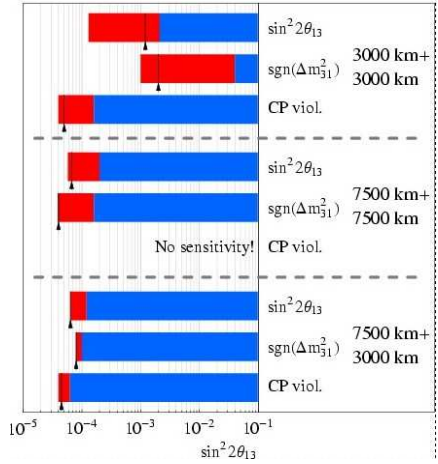

*Figure 16 The sensitivity reaches as functions of $\sin^2 2\theta_{13}$ for $\sin^2 2\theta_{13}$ itself, the neutrino mass hierarchy, and maximal CP violation for each of the indicated baseline combinations. The bars show the ranges in $\sin^2 2\theta_{13}$ where sensitivity to the corresponding quantity can be achieved at the $3\sigma$ CL. The dark (red) bars show the variation in the result as $\Delta m^2_{21}$ is varied within its present uncertainty. Figure from [Huber03].*

The above studies were made only using the "golden channel", and for very high muon momentum (cut at 4-5 GeV). A more granular detector may be able to either detect electrons, perhaps even measure their charge ("platinum" channel) or detect tau leptons and measure their charge ("silver" channel). In the studies made assuming a large liquid argon detector [Bueno00], the detected events could be classified in four classes: charged-current electrons; right-sign muons, wrong-sign muons, events with no leptons. An example of the set of energy spectra for these classes, for positive and negative muons circulating in the ring, is given in Figure 17.

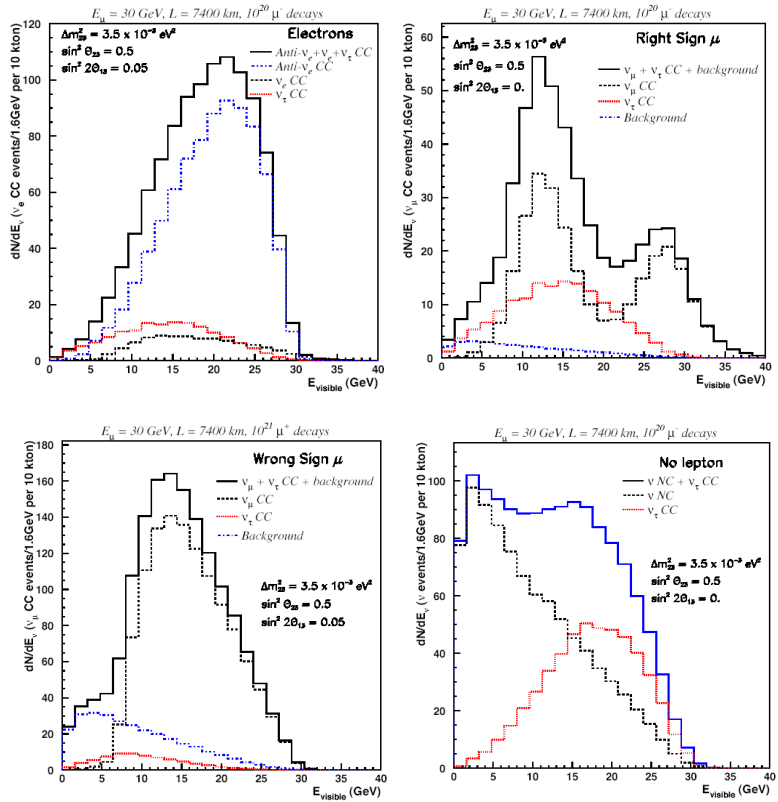

*Figure 17 : Four classes of events studied in a liquid argon TPC with muon charge identification. From the top, left to right: events with high-energy electrons, right-sign muons, wrong-sign muons, no charged leptons [Bueno00]. The baseline here is 7200 km.*

Performing measurements at potential Neutrino Factories will face the existence of correlations and degeneracies in parameter space [Cervera00],[Burguet01],[Minakata01][ not found in references],[Barger02] which make the simultaneous determination of all the unknowns rather difficult. The importance of having good neutrino energy resolution or combining the measurements of the *golden* oscillation probabilities at several experiments with different

< $E_\nu$ /L > (or different matter effects) have been proposed to overcome this problem [Burguet01],[Barger02],[Burguet02].

In addition, the measurement of the *silver* channels [Donini02] $\nu_e \to \nu_\tau$, $\bar{\nu}_e \to \bar{\nu}_\tau$ besides the *golden* one, although it is experimentally more challenging, is extremely powerful in reducing these correlations. The *silver* channel also provides a test of unitarity of the Neutrino mixing matrix! In fact, it has been shown that while the combination of beta-beam and Superbeam could not help in solving the degeneracies, the combination of one of them with the Neutrino Factory Golden and Silver channel can, instead, be used to solve completely the eightfold degeneracy. It is even advocated, although full demonstration is needed, that with a Neutrino Factory with two baselines and detectors able to measure both the golden and silver channels in addition to the disappearance channels, a fully unambiguous determination of oscillation parameters could be achieved.

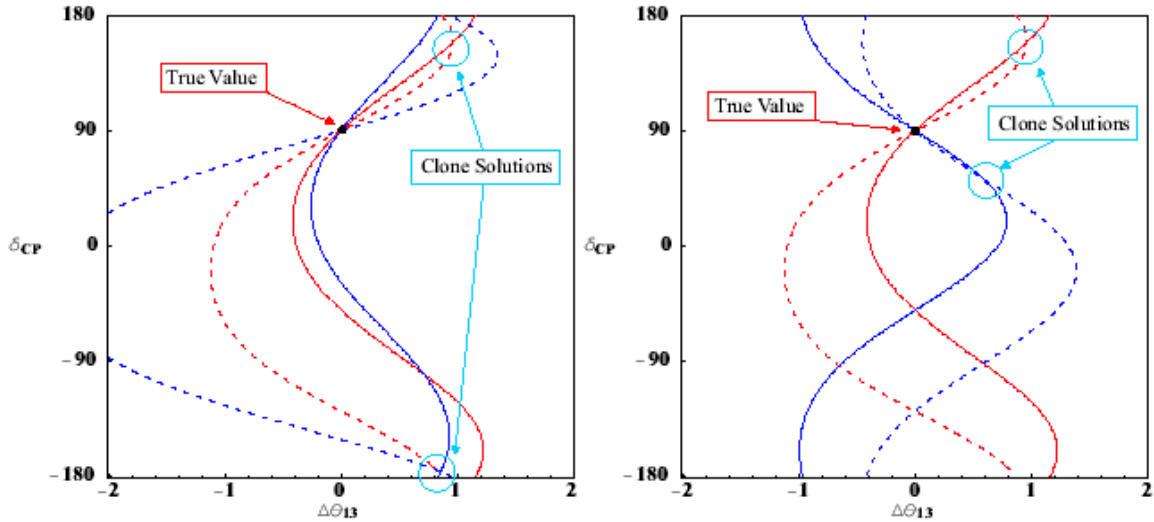

*Figure 18 Solving degeneracies (from [Rigolin04]). The parameter space shown is the variation {Δθ$_{13}$, δ$_{CP}$} around the true solution in the {θ$_{13}$, δ$_{CP}$} plane. The lines show the locus where the same number of events would be observed. Full lines, neutrino exposure; dashed lines antineutrino exposure. On the left, the red and blue lines show two different base lines (730 and 3500 km) while on the right the red and blue lines show the golden and silver channel.*

These studies were done for the Neutrino Factory some years ago for a muon energy of $E_\mu$ =20-50 GeV and a baseline for the *golden* measurement of a few thousand kilometres [Cervera00]. The combination of this measurement, using a 40 Kton iron calorimeter [Cervera00a], plus the *silver* one in an Opera-like tau-neutrino detector [Donini02] results in a great physics potential. The sensitivity to $\sin^2\theta_{13}$ is below $10^{-4}$ and there is a 99% CL discovery potential for CP violation if δ> $10^0$. In addition, the atmospheric parameters can be determined with a 1% precision and the sign of $\Delta m^2_{23}$ can be measured in a large range of parameter space. These studies will need to be actualised to take into account progress both in accelerator and detector design.

## 2.6  High-precision neutrino scattering

As discussed in [Mangano01], [Bigi01], the neutrino beams at the end of the straight section of a Neutrino Factory offer an improvement in flux by several orders of magnitude over conventional beams, allowing several times $10^8$ events to be collected per kilogram and per year (Figure 19). This could allow a new generation of neutrino experiments, with detailed studies of nucleon structure, nuclear effects, spin structure functions and final state exclusive processes. Precision tests of the Standard Model could be carried out in neutrino scattering on nucleon or electron target, as well as a precise determination of neutrino cross-sections and flux monitoring with permil accuracy.

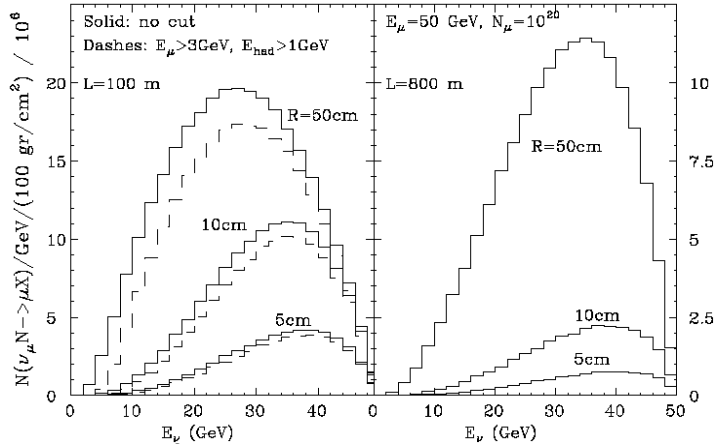

Figure 19 Event rates at the exit of the straight sections in a Neutrino Factory. Note the scale: $10^4$ evts per gram per year per bin.

## 2.7 Muon physics

A high-intensity proton source could certainly produce many low-energy muons [Aysto01] and thus, provided the beam and experiments can be designed to do so, provide opportunities to explore rare decays, such as $\mu \rightarrow e \gamma$, $\mu \rightarrow e e e$, or the muon conversion $\mu N \rightarrow e N$, which are lepton-number-violating processes. While the See-Saw mechanism provides a very appealing explanation of neutrino masses and mixings, its inclusion in Supersymmetric models almost invariably leads to predictions in excess or close to the present limits for these processes.

It is therefore quite possible that one of these processes will be discovered in the upcoming generation of experiments (MEG [MEG] at PSI, MECO at BNL) in which case a detailed study would become mandatory. If not, further search with higher sensitivity would be in demand. It should be emphasized that the three processes are actually sensitive to different parameters of these models, and thus complementary from both the experimental and theoretical points of view.

Another fundamental search would clearly be the search for a muon electric dipole moment (EDM), which would require modulation of a transverse electric field for muons situated already at the magic velocity where the magnetic precession and the anomalous (g-2) precession mutually cancel.

## 2.8 Muon Colliders

Finally, it is worth keeping in mind that the Neutrino Factory is the first step towards muon colliders. As shown in [muoncollider], the relevant characteristics of muons are that, compared to electrons, i) they have a much better defined energy, since they hardly undergo synchrotron radiation or beamstrahlung, ii) their coupling to the Higgs bosons is multiplied by the ratio $(m\mu/me)^2$, thus allowing s-channel production with a useful rate.

These remarkable properties make muon colliders superb tools for the study of Higgs resonances, especially if, as predicted in supersymmetry, there exists a pair H, A of opposite CP quantum numbers which are nearly degenerate in mass. The study of this system is extremely difficult with any other machine and a unique investigation of the possible CP violation in the Higgs system would become possible.

# 3 Comparison of facilities and open questions

The performances of various superbeams, betabeams and Neutrino Factories have been compared extensively with varying methods and presentations. One set of comparative plots were prepared for the MMW workshop, and are shown in Figure 20. Another set of plots could be seen at NUFACT05 and are shown in Figure 21. The comparison of these curves points to a number of questions and comments.

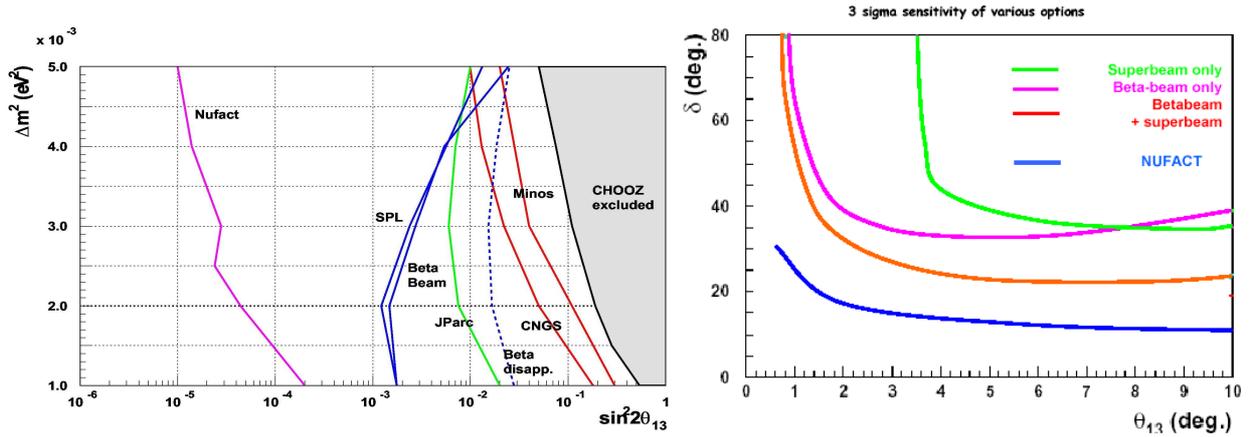

*Figure 20 Relative sensitivity to $\theta_{13}$ (top) and to the CP violating phase $\delta$ (bottom), of various options for future neutrino facilities based on a high intensity proton driver – from the 2004 MMW workshop.*

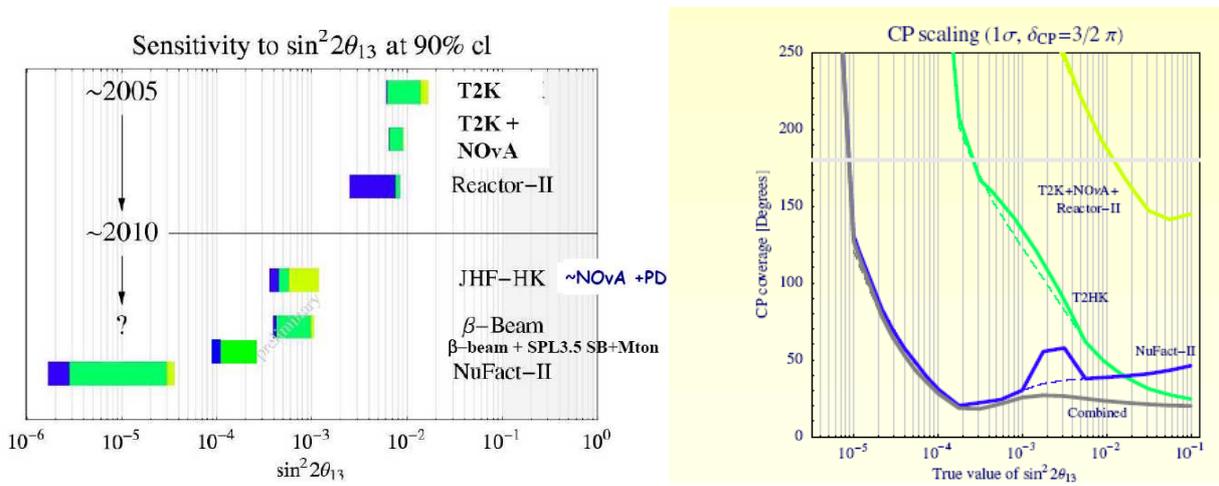

*Figure 21 Comparison of different options for future neutrino facilities. On the left, the sensitivity to $\sin^2 2\theta_{13}$ described with the convention of the MTU group [Lindner02], where: the intrinsic statistical sensitivity of the experiment can be readout on the left of the band corresponding to each considered experiment (this assumes that all other oscillation parameters are infinitely and unambiguously well known); the experimental systematics are shown next (including those from matter effects) as a blue band; the effect of other parameter uncertainties is shown by the green bad, the effect of ambiguities in case of multiple solutions in parameter space are shown as a yellow band. On the right, precision in the measurement of the CP phase $\delta$, expressed as the size of the 90% CL spot in parameter space (for small 'coverage' this would be about three times the error).*

It is clear that from the first set of plots (Figure 20) one would conclude that a neutrino factory is the most powerful device to measure CP violation, determine the mass hierarchy and extract the best precision on $\sin^2 2\theta_{13}$, regardless of the value of this mixing angle. If this is correct, then one would conclude that there is no need to wait for the next generation of experiments to go forward with it.

On the contrary, the second set of plots seems to indicate that the superiority of the neutrino factory is a $\sin^2 2\theta_{13}$ – dependent statement and that it is "urgent to wait". At this point one could argue that the breaking point is around $\sin^2 2\theta_{13} \sim 0.01$ and that the combination of MINOS at NUMI, OPERA at CNGS , the next reactor experiment and the first runs of T2K should give already a clear indication of where this parameter lies – by 2010 or so (Figure 5) and that one should get prepared in any case. Nevertheless, a closer investigation of these conclusions is necessary and instructive.

As M. Mezzetto pointed out in his NUFACT05 presentation [Mezzetto05], and for the reason advocated earlier (Figure 4), measuring the CP violation for large values of $\theta_{13}$ is not too difficult from the point of view of statistics, but is extremely challenging from the point of view of systematic errors, because the asymmetry is small. As pointed out earlier, one solution is to go for the second maximum, where the effect is larger – but the statistics much smaller. In addition, the interplay between matter effects and CP violation requires either a good knowledge of matter effects from external information, or a more sophisticated analysis to extract them from the oscillation data themselves as it has been done in [Bueno00]. The next generation of studies should clearly address the issues of systematic errors in a systematic and, as always when dealing with systematics, creative way.

Putting together the information gathered at the time of the NUFACT05 workshop, a more complete comparative picture has been drawn in Figure 22, emphasizing the role of systematic errors. These plots were produced by P. Huber with GLOBES [Globes] and show the sensitivity to CP violation at $3\sigma$ CL ($\Delta\chi^2$=2.9): sensitivity to CP violation is defined, for a given point in the $\theta_{13}$-$\delta$ plane, by being able to exclude $\delta$=0 and $\delta$=$\pi$. Degeneracies and correlations are fully taken into account. For all set-ups the appropriate disappearance channels have been included. The beta beam is lacking muon neutrino disappearance, but the result does not change if T2K disappearance information is included in the analysis. In all cases systematics between neutrinos, anti-neutrinos, appearance and disappearance is uncorrelated. For all set-ups with a water Cherenkov detector the systematics applies both to background and signal, uncorrelated.

The neutrino factory assumes $3.1 \cdot 10^{20}$ $\mu^+$ decays per year for 10 years and $3.1 \cdot 10^{20}$ $\mu^-$ decays per year for 10 years. It has one 100 kton detector at 3000 km and another 30 kton detector with 30 kton at 7000 km. The density errors between the two baselines are uncorrelated. The systematics are 0.1% on the signal and 20% on the background, uncorrelated. The detector threshold and the other parameters are taken from [Huber02], and follow closely [Cervera00a].

The beta beam assumes $5.8 \cdot 10^{18}$ He decays per year for five years and $2.2 \cdot 10^{18}$ Ne decays per year for five years. The detector mass is 500 kton. The detector description and the glb-file is from [Mezzetto05]. The SPL set-up is taken from [Campagne05], and the detector mass is 500 kton. The systematic errors on signal efficiency (or cross-sections) and background are 2% or 5%.

The T2HK set-up is taken from [Huber02] and closely follows the LOI [Itow01]. The detector mass is 1000 kton and it runs with 4MW beam power, 6 years with anti-neutrinos and 2 years with neutrinos. The systematic error on both background and signal is 5%.

The oscillation parameters were [Maltoni],[Lisi]: $\Delta m^2_{13}$ = 0.0024 eV$^2$, $\Delta m^2_{12}$=0.00079 eV$^2$, $\theta_{23}$=$\pi/4$, $\theta_{12}$=0.578. The input errors are (at 1 sigma): 10% on $\Delta m^2_{13}$, $\theta_{23}$, $\theta_{12}$, 4% on $\Delta m^2_{12}$ and 2 or 5% on the matter effect $\rho$. For the Oulu mine site in Finland [Oulu], a description of the baseline and the related matter uncertainties is available [Oulu-matter], indicating that the 2% systematic error could possibly be achieved, but the introduction of the details of this information into Globes is still to be made.

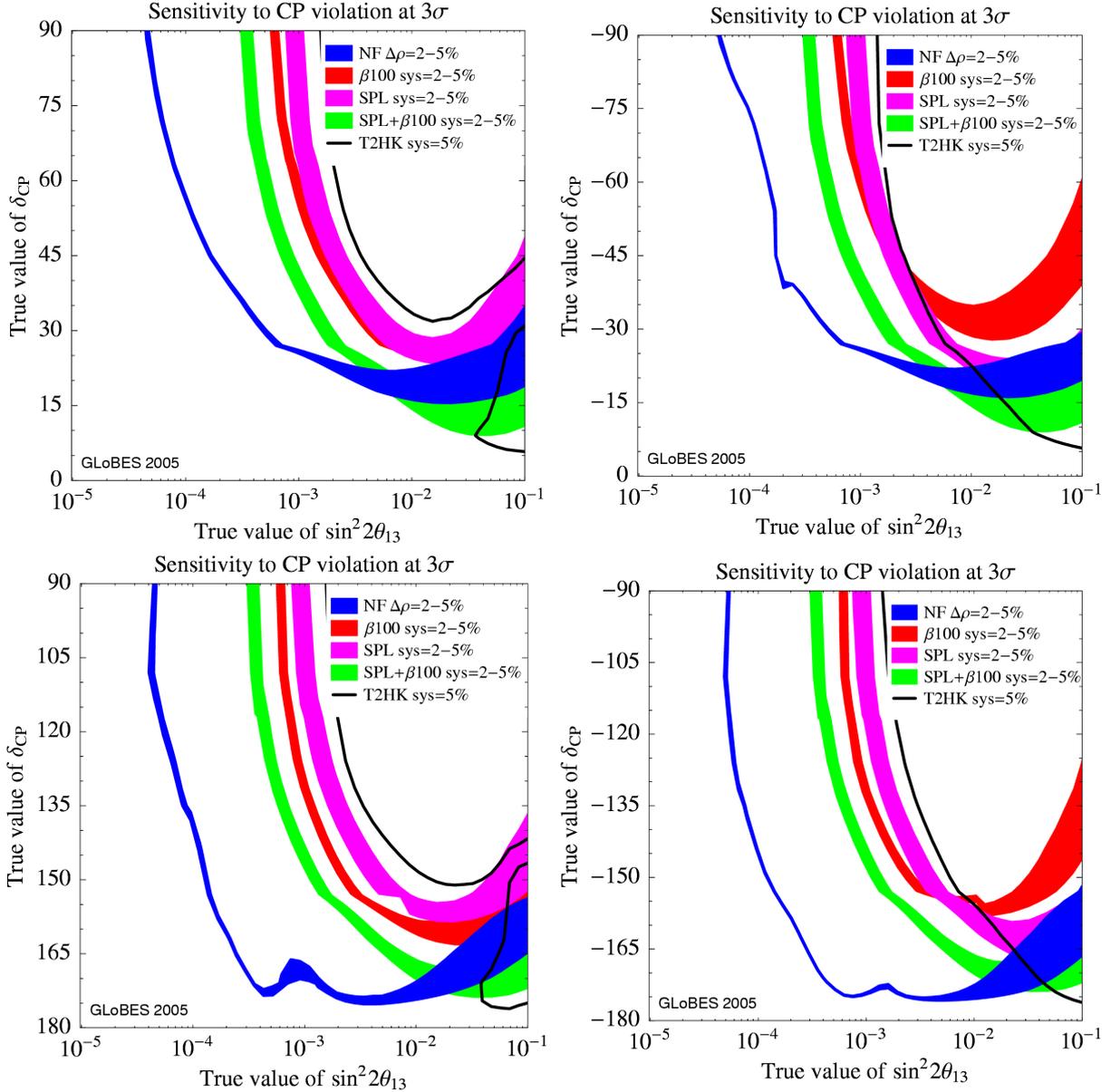

*Figure 22 Provisional estimates of the 3σ sensitivities of various proposed neutrino facilities to the CP-violating phase δ in the simplest three-generation model of neutrino mixing (plots are drawn such that the accessible range is always above the lines). The black line assumes that JPARC delivers 4 MW for 8 years to Hyper-Kamiokande, a megaton water Cherenkov detector. The magenta line is the SPL-based superbeam running for 10 years aimed at a 440 kton water Cherenkov detector at a distance of 130 km; the red line is the γ=100 β-beam aimed at the same detector; and the green line is a combination of the two. The blue line is the neutrino factory optimised for small values of $\theta_{13}$, aiming at detectors 3000 and 7000 km away. The thickness of the lines shows the effect of varying the systematic errors related to cross sections or matter effects within the indicated ranges. A definitive version of this comparison is a deliverable of the International Scoping Study.*

The discussion necessary to establish reasonable systematic errors in measuring the CP or T asymmetry should include the following questions:

1. How does one measure the cross-section*efficiency of the appearance channel in a beam with only one flavour? (superbeam or beta-beam alone)

2. What kind of near detector will be needed? It seems impossible to talk about flux systematics without at least a straw man concept for the near detector.

3. What is the energy sensitivity of these issues – which should be quite serious at low energies (E ~ few 100 MeV) and gradually become easier at higher energies. The Neutrino Factory provides all channels in the same beam line/detector, but can one make use of this?

In order to go further, one must address a set of specific remarks and questions to the various types of facilities

## 3.1 Neutrino Factory issues.

One must remember that most of the Neutrino Factory studies have been made in the years 1998-2001, and would highly benefit from a refreshed look. There is no ordering in the following questions, and there should probably be some reflection on understanding which one should be tackled first.

1. Since the early part of the studies it has been advocated that having two different locations allows to decipher the matter effect from the CP violation, as they have different distance dependence. With the realization that the energy distribution could also resolve this issue, do we really need far locations at two different distances?

2. The detector that was simulated (a large magnetized iron calorimeter) and the analysis that was performed in the studies so far have advocated a muon momentum cut at about 4-5 GeV, in order to get a very clean and background free sample of wrong-sign muons. This is clearly favourable to obtaining a very good sensitivity to $\theta_{13}$, but leads to a poor efficiency at low momentum. Perhaps as a consequence, the optimal distance was found to be about 3000 km, for which the first oscillation maximum is located at about 6 GeV. The second oscillation maximum at 2 GeV is then completely lost. Can this be improved, by a more finely segmented iron calorimeter, or a magnetized fine-grained detector (fully active scintillator or Liquid Argon detector imbedded in a magnetic field) ?

3. Can one really eliminate all degeneracies by combination of energy distribution and analysis of different channels (tau, muon, electron, both signs, NC…) and what is the need of two baselines to do so?

4. What are the systematics on flux control? (The CERN yellow report claims that a precision of $10^{-3}$ should be achievable)

5. What are the a priori systematics on the knowledge of the matter effect? How can this be determined from the data, and what are the conditions on the experiment to be able to do so?

6. What is the best baseline and consequently optimal stored muon energy? This is an important question as this is a cost-driver in the accelerator (in study II the cost of neutrino factory was given as ~ 1500M\$ + 400M\$*E/20GeV)

## 3.2 Superbeam+beta-beam issues

1. The first question of course is that of finding the energy and distance combination that allows both beta-beam and superbeam to be used optimally. The difficulty with a low energy scenario such as CERN-Fréjus is that the short baseline imposes an oscillation maximum at $E_\nu$ ~250 MeV. This is barely above the muon Cherenkov threshold, and commensurate with both the Fermi momentum and the muon mass, thus making it very sensitive to nuclear effects, muon mass correction etc… Before being considered seriously, this experiment must be validated by a convincing study of systematic errors and of the means to control them. A strategy of ancillary measurements in the near detector station could be the answer to this issue.

2. One of the arguments in favour of the beta-beam is that the production targets necessary for ion production are efficient enough not to require very high primary proton power. This advantage is reduced if a 4MW superbeam is necessary to master systematic errors. The question is then: what is the importance of the superbeam in this scheme? Is it only the beauty of T violation or much increased sensitivity – or the need to have a (known) source of muon neutrinos to calibrate the cross-sections and efficiencies of the detector to the $\nu_e \to \nu_\mu$ "golden" signal?

3. At which neutrino energy can one begin to use the event energy distribution, which is a very powerful cross-check of the experiment? Here again, the effect of Fermi motion appears, in that it affects the ability to reconstruct the event kinematics. Here again, what is the impact of muon Cherenkov threshold?

4. Should energy remain an adjustable parameter once the long baseline distance has been chosen? One could imagine that this could be achieved by moveable horns for the Wide Band Beam, by varying the beam axis for the off-axis beam, or by changing the energy of the storage ring in the beta beam.

5. What is the relationship between beta-beam and superbeam energies, vs intensity?6. What is really the cost of the detector? For the water Cherenkov, what PM coverage is needed as function of neutrino energy? For the Liquid Argon or scintillator alternatives, can they, thanks to better granularity and acceptance, compete with the water Cherenkov given the unavoidable limitations in detector mass?

# 4 Conclusions

Neutrinos have unveiled for us a new layer of reality beyond the Standard Model. It is possible that the answer to their extraordinary small masses is trivial, or it may have its origin at a much higher scale that what will be investigated with the next generations of accelerators (LHC, ILC, CLIC). It may be that CP/T violation will be observed and it may be that neutrinos have a different concept of hierarchy than charged fermions. These questions are simple but fundamental. They justify a considerable investment in time and resources from the community. There exists an opportunity for a powerful facility to be decided together and built in the next decade, and the question is: which one? The scoping study will not answer this question but should identify the work that will be needed to make this decision in the best possible way. I hope this short reflection and list of questions will prove useful.

## Acknowledgements

This presentation and article summarize the work of several hundred of members of the CERN and ECFA studies of a Neutrino Factory Complex, which was described in the very complete [ECFAreport], none of which would have been possible without the pioneering work of the Neutrino Factory and Muon Collider collaboration [MuColl]. We also acknowledge the support of the European Community-Research Infrastructure Activity under the FP6 "Structuring the European Research Area" programme (CARE, contract number RII3-CT-2003-506395)

# 5 References

| | |
|---|---|
| [Apollonio02] | M. Apollonio et al., 'Oscillation Physics with a Neutrino Factory' arXiv:hep-ph/0210192 in "ECFA/CERN Studies of a European Neutrino Factory Complex" Blondel, A (ed.) et al.CERN-2004-002.- ECFA-04-230, p85 |
| [APS04] | "Neutrino Factory and Beta Beam Experiments and Developments", (Eds. S. Geer and M. Zisman), Report of the Neutrino Factory and Beta Beam Working Group, APS Multi-Divisional Study of the Physics of Neutrinos, July 2004. |
| [Atmos] | Y. Fukuda et al. (Super-Kamiokande Collaboration), Phys. Rev. Lett. 81 (1998) 1562, hep-ex/9807003;see also M. C. Sanchez et al. Soudan 2 Collaboration, Phys. Rev. D 68 (2003) 113004 hep-ex/0307069; M. Ambrosio et al. MACRO Collaboration, Phys. Lett. B 566 (2003) 35 hepex/0304037; Y. Ashie et al. Super-Kamiokande Collaboration, Phys. Rev. Lett. 93 (2004) 101801 hep-ex/0404034. |
| [Aut99] | B. Autin, A. Blondel and J. Ellis eds, CERN yellow report CERN 99-02, ECFA 99-197 |
| [Aysto01] | J. Aysto et al., 'Physics with Low-Energy Muons at a Neutrino Factory Complex', CERN-TH/2001-231, in 'ECFA/CERN Studies of a European Neutrino Factory Complex ' Blondel, A (ed.) et al.CERN-2004-002.- ECFA-04-230, p259. |
| [Barbieri94] | R. Barbieri and L.J. Hall Phys. Lett. B338 (1994) 212 |
| [Barger02] | V. Barger, S. Geer, R. Raja, K. Whisnant, Phys. Rev. D63 (2001) 033002. |
| [Barger02a] | V. Barger, D. Marfatia and K. Whisnant, 'Breaking eight-fold degeneracies in neutrino CP violation, mixing, and mass hierarchy' Phys. Rev. D65 (2002) 073023, [arXiv:hep-ph/0112119]. |
| [BENE] | Beams for European Neutrino Experiments is a Networking Activity (http://bene.na.infn.it/) supported by the EC and most major European Agencies in the framework of the FP6 Integrated Activity CARE (Coordinated Accelerator R&D in Europe) http://esgard.lal.in2p3.fr/Project/Activities/Current/. The scope of BENE is also described in more detail in the following document: http://esgard.lal.in2p3.fr/Project/Activities/Current/Networking/N3/BENE-downsized-11.doc |
| [BENE05] | BENE interim report to appear as CERN yellow report 2006-xx |
| [Bernabeu04] | J. Bernabeu, J. Burguet-Castell, C. Espinoza and M. Lindroos, Monochromatic neutrino beams, hep-ph/0505054 |
| [beta-beam] | http://beta-beam.web.cern.ch/beta-beam/ |
| [Bigi01] | I. I. Y. Bigi et al., 'The Potential for Neutrino Physics at Muon Colliders and Dedicated High Current Muon Storage Rings' Phys. Rept. 371 (2002) 151, hep-ph/0106177. |
| [Blondel04] | A. Blondel 'Neutrino Factory scenarios' presentation at NUFACT04, http://www-kuno.phys.sci.osaka-u.ac.jp/~nufact04/. |
| [Bueno00] | A. Bueno, M. Campanelli, A. Rubbia, 'Physics potential at a neutrino factory: can we benefit from more than just detecting muons?' Nucl.Phys. B589 (2000) 577-608, hep-ph/0005007. |
| [Burguet01] | J. Burguet-Castell et al, Nucl. Phys. B608 (2001) 301, hep-ph/0103258. |
| [Burguet02] | J. Burguet-Castell et al, Nucl. Phys. B646 (2002) 301, hep-ph/0207080. |
| [Burguet03] | J. Burguet-Castell, D. Casper, J.J. Gómez-Cadenas, P. Hernández, F. Sanchez, Nucl.Phys. B695 (2004) 217-240, hep-ph/0312068. |
| [Burguet05] | |
| [Campagne05] | J.E. Campagne, hep-ex/0511013. |
| [CARE] | FP6 Integrated Activity (Coordinated Accelerator R&D in Europe) http://esgard.lal.in2p3.fr/Project/Activities/Current/. |
| [Cervera00] | A. Cervera et al, Nucl. Phys. B579 (2000) 17; A. Cervera, Invited Talk at Nufact04, the Int. Workshop on Neutrino Factories and Superbeams, Osaka, July 2004, http://www-kuno.phys.sci.osaka-u.ac.jp/~nufact04/ |
| [Cervera00a] | A. Cervera, F. Dydak and J.J. Gomez-Cadenas, Nucl. Instr. Meth. A 451 (2000) 123. |
| [CHOOZ] | M. Apollonio et al. CHOOZ Collaboration, Eur. Phys. J. C 27 (2003) 331, hep-ex/0301017. |
| [CUORE] | R. Ardito et al., hep-ex/0501010. |
| [Cuoricino] | C. Arnaboldi et al., Phys. Rev. Lett. 95 (2005) 142501, hep-ex/0501034. |
| [Donini02] | A. Donini, D. Meloni, P. Migliozzi, Nucl. Phys. B646 (2002) 321, hep-ph/0206034; D. Autiero et al., 'The synergy of the golden and silver channels at the Neutrino Factory' hep-ph/0305185. |
| [ECFAreport] | 'ECFA/CERN studies of a European Neutrino Factory complex', A. Blondel et al, ed CERN 2004-002 (2004) http://preprints.cern.ch/cernrep/2004/2004-002/2004-002.html |
| [Gilardoni] | S. Gilardoni, PhD thesis University of Geneva (2004). See also Neutrino Factory notes 4, 38, 42, 77, 80, 81, 86, 126, 129, 134,138 accessible from http://slap.web.cern.ch/slap/NuFact/NuFact/NFNotes.html |
| [Globes] | P. Huber, M. Lindner, W. Winter, Comput. Phys. Commun. 167 (2005) 195, hep-ph/0407333. |
| [GNO] | M. Altmann et al. GNO Collaboration, Phys. Lett. B 490 (2000) 16,hep-ex/0006034. |
| [Gru02] | P. Gruber et al, 'The Study of a European Neutrino Factory Complex', Neutrino Factory Note 103(2002) CERN/PS/2002-080(PP), in ''ECFA/CERN Studies of a European Neutrino Factory Complex'' Blondel, A (ed.) et al.CERN-2004 002.- ECFA-04-230, p7 |
| [HM-KK] | H. V. Klapdor-Kleingrothaus, A. Dietz and I. V. Krivosheina, Found. Phys. 32 (2002) 1181 Erratum-ibid. 33 (2003) 679, hep-ph/0302248. |
| [Homestake] | B. T. Cleveland et al., Astrophys. J. 496 (1998) 505. |
| [Huber02] | P. Huber, M. Lindner and W. Winter, Nucl. Phys. B645 (2002) 3, hep-ph/0204352. |
| [Huber02] | P. Huber, M. Lindner, W. Winter, Nucl.Phys. B645 (2002) 3-48, hep-ph/0204352. |
| [Huber03] | P. Huber and W. Winter; Phys. Rev. D68 (2003) 037301; hep/ph-0301257. |
| [IGEX1] | Y. G. Zdesenko, F. A. Danevich and V. I. Tretyak, Phys. Lett. B 546 (2002) 206. |
| [IGEX2] | A. M. Bakalyarov, A. Y. Balysh, S. T. Belyaev, V. I. Lebedev and S. V. Zhukov C03-06-23.1 Collaboration, Phys. Part. Nucl. Lett. 2 (2005) 77; Pisma Fiz. Elem. Chast. Atom.Yadra 2 (2005) 21, hep-ex/0309016. |
| [Itow01] | Y. Itow, et al., hep-ex/0106019. |
| [Japnufact] | ''A Feasibility Study of A Neutrino Factory in Japan'', Y. Kuno, ed., http://www-prism.kek.jp/nufactj/index.html |
| [K2K] | E. Aliu et al. K2K Collaboration, Phys. Rev. Lett. 94 (2005) 081802 hep-ex/0411038]. |
| [KamLAND02] | K. Eguchi et al., KamLAND Coll., First Results from KamLAND: Evidence for Reactor Anti-Neutrino Disappearance, Phys. Rev. Lett. 90 (2003) 021802 -- hep-ex/0212021. |
| [KamLAND04] | KamLAND collaboration, Measurement of neutrino oscillations with KamLAND: Evidence of spectral distortion, hep-ex/040621. |
| [KATRIN] | KATRIN Coll., hep-ex/0109033. |
| [LEPEW] | Precision Electroweak Measurements on the Z Resonance. The LEP and SLD collaborations, hep-ex/0509008, to appear as physics report. [Leptogenesis] M. Fukugita and T. Yanagida, Phys. Lett. B174 (1986) 45; W. Buchmuller, P. Di Bari and M. Plumacher, Nucl. Phys. B665 (2003) 445; G. F. Giudice, A. Notari, A. Riotto and A. Strumia, Nucl. Phys. B685 (2004) 89; W. Buchmuller, P. Di Bari and M. Plumacher, Ann. Phys. 315 (2005) 303. |
| [Lindner02] | M. Lindner, 'The physics potential of future long baseline neutrino oscillation experiments', to appear in ''Neutrino Mass'', Springer Tracts in Modern Physics, ed. G. Altarelli and K. Winter, hep-ph/0209083 |
| [Lisi] | E. Lisi, 'Neutrino mases and mixing', NUFACT05; G. L. Fogli, E. Lisi, A. Marrone and A. Palazzo, hep-ph/0506083. |
| [Lombardi] | A. Lombardi, 'A 40-80 MHz System for Phase Rotation and Cooling', Nufact Note 34. See also notes 41, 102, 119. http://slap.web.cern.ch/slap/NuFact/NuFact/NFNotes.html |
| [LSND] | A. Aguilar et al. LSND Collaboration, Phys. Rev. D 64 (2001) 112007, hep-ex/0104049. |
| [Mainz] | C. Weinheimer, Nucl. Phys. Proc. Suppl. 118 (2003) 279. |
| [Maltoni] | M. Maltoni, T. Schwetz, M.A. Tortola, J.W.F. Valle, New J.Phys.6 (2004) 122, hep-ph/0405172. |
| [Mangano01] | M. L. Mangano et al., 'Physics at the front-end of a Neutrino Factory: a quantitative appraisal',hep-ph/0105155, in 'ECFA/CERN Studies of a European Neutrino Factory Complex ' Blondel, A (ed.) et al.CERN-2004-002.- ECFA-04-230, p187 |
| [MARE] | A. Monfardini et al., hep-ex/0509038. |
| [MEG] | The MEG experiment: search for the m→eg at PSI, September 2002, available at http://meg.psi.ch/docs/prop_infn/nproposal.pdf |
| [Mezzetto03] | M. Mezzetto, J. Phys. G.29 (2003) 1771 and 1781. |
| [Mezzetto05] | M. Mezzetto, hep-ex/0511005. |
| [MICE] | The International Muon Ionization Experiment MICE, http://hep04.phys.iit.edu/cooldemo |
| [Minakata01] | H. Minakata and H. Nunokawa, JHEP 0110 (2001) 001 [arXiv:hep-ph/0108085]. |
| [MiniBoone] | E. Church et al. BooNe Collaboration, nucl-ex/970601. |
| [MMW] | Workshop on Physics at a Multi Megawatt Proton driver, CERN-SPSC-2004-SPSC-M-722, http://physicsatmwatt.web.cern.ch/physicsatmwatt/Contributions/workshop-summary.pdf |
| [MSW] | L. Wolfenstein, Phys. Rev. D17 (1978) 2369. S.P. Mikheev and A.Y. Smirnov, Nuovo Cim. C9 (1986) 17. |
| [MuColl] | The Muon Collider and Neutrino Factory Collaboration, see the web site http://www.cap.bnl.gov/mumu/ which contains also references to several physics studies. |
| [Mucool] | The Muon Cooling Experimental R&D http://www.fnal.gov/projects/muon_collider/cool/cool.html |
| [muoncollider] | See for instance S. Kramlet al, ,Physics opportunities at $\mu^+\mu^-$ Higgs factories, in [ECFAreport], p337. |
| [Nelson] | J. Nelson, presentation at NUFACT05, http://www.lnf.infn.it/conference/2005/nufact05/ and at the ISS meeting at CERN http://dpnc.unige.ch/users/blondel/ISSatCERN.htm |
| [NEMO3] | R. Arnold et al. NEMO Collaboration, hep-ex/0507083. |
| [NOvA] | NovA: Proposal to Build a 30 Kiloton Off-Axis Detector to Study Neutrino Oscillations in the Fermilab NuMI Beamline, hep-ex/0503053. |
| [nufact] | S. Geer, Phys. Rev. D57 (1998) 6989; A. De Rújula, M.B. Gavela and P. Hernández, Nucl. Phys. B547 (1999) 21; A. Blondel et al., Nucl. Instrum. Methods Phys. Res. A 451 (2000) 102; For recent reviews, see M. Apollonio, et al, in CERN-04/02, (2004) hep-ex/0210192; J. J. Gómez-Cadenas and D.A. Harris, 'Physics opportunities at neutrino factories' Ann. Rev. Nucl. Part. Sci. 52 (2002) 253 and the annual proceedings of the International Nufact Workshop. |
| [NuFactPhys] | B. Autin, A. Blondel and J. Ellis eds, CERN yellow report CERN 99-02, ECFA 99-197; C. M. Ankenbrandt et al., Phys. Rev. ST Accel. Beams 2, 081001 (1999); A. Blondel et al., Nucl. Instrum. Methods Phys. Res., A 451 (2000) 102; C. Albright et al., FERMILAB-FN-692, hep-ex/0008064; D. Harris et al., Snowmass 2001 Summary, hep-ph/0111030; A. Cervera et al., Nucl. Phys. B579, 17 (2000), Erratum-ibid.B593:731-732,2001; M. Koike and J. Sato, Phys. Rev. D62 (2000) 073006. |
| [Oulu] | University of Oulu, CUPP project, http://cupp.oulu.fi/ |
| [Oulu-matter] | E. Kozlovskaya, J. Peltoniemi, J. Sarkamo, *'The density distribution in the Earth along the CERN-Pyhäsalmi baseline and its effect on neutrino oscillations',* CUPP-07/2003 |
| [Plunkett] | R. Plunkett, 'The Fermilab neutrino program', http://www.lnf.infn.it/conference/nufact05/talks/Plenary/Plunkett_Plenary.pdf |
| [Rigolin04] | S. Rigolin, 'Why care about ($\theta_{13},\delta$) degeneracy at future neutrino experiments', Rencontres de Moriond 2004, to appear in the proceedings, hep-ph/0407009 |
| [Rubbia] | A. Rubbia, Neutrino detectors for future experiments, A.Rubbia, Nucl. Phys. B (Proc. Suppl.) 147 (2005) 103; A. Rubbia, these proceedings and references therein. |
| [Ruj99] | A. De Rujula, M.B. Gavela, P. Hernandez, Nucl. Phys. B547 (1999) 21, hep-ph/9811390. |
| [Sage] | J. N. Abdurashitov et al. SAGE Collaboration, J. Exp. Theor. Phys. 95 (2002) 181; Zh.Eksp. Teor. Fiz. 122 (2002) 211, astro-ph/0204245. |
| [SNO01] | Q. R. Ahmad et al., SNO Coll, Phys. Rev. Lett. 87, 071301 (2001) |
| [SNO02] | Q. R. Ahmad et al., SNO Coll., nucl-ex/0204008. |
| [StudyI] | Feasibility Study on a Neutrino Source Based on a Muon Storage Ring, D.Finley, N.Holtkamp, eds. (2000), http://www.fnal.gov/projects/muon_collider/reports.html |
| [StudyII] | 'Feasibility Study-II of a Muon-Based Neutrino Source', S. Ozaki, R. Palmer, M. Zisman, and J. Gallardo, eds. BNL-52623, June 2001, available at http://www.cap.bnl.gov/mumu/studyii/FS2-report.html ; M.M. Alsharo'a et al., Phys. Rev. ST Accel. Beams 6, 081001 (2003). |
| [Super-Ksolar] | S. Fukuda et al. Super-Kamiokande Collaboration, Phys. Lett. B 539 (2002) 179 |
| [target-exp] | R.J. Bennett et al, 'Studies of a target system for a 4MW 24 GeV proton beam', CERN-INTC proposal 2003-033, April 2004. |
| [Troitsk] | V. M. Lobashev et al., Nucl. Phys. Proc. Suppl. 91 (2001) 280. V. M. Lobashev, proceeding of "Neutrino Telescopes 2005", Venice, 507-517 |
| [Zucchelli] | P. Zucchelli, 'A novel concept for a neutrino factory: the beta-beam', Phys. Let. B, 532 (2002) 166-172 |